\documentclass[twocolumn,amsmath,amssymb,nofootinbib,tighten,floatfix,prb]{revtex4}

\usepackage{graphicx}
\usepackage{dcolumn} 
\usepackage{bm, bbm}      
\usepackage{curves}
\usepackage{epic}
\usepackage{wasysym}
\usepackage{epsf, times}
\usepackage{subfigure}
\usepackage{eufrak}
\usepackage{amsthm}

\newcommand\bftheta{\boldsymbol{\theta}}

\begin{document}

\title{
A quantum topological phase transition at the microscopic level
      }

\author{
Claudio Castelnovo$^1$ 
and 
Claudio Chamon$^2$
       }
\affiliation{
$^1$ 
Rudolf Peierls Centre for Theoretical Physics, 
University of Oxford, Oxford, OX1 3NP, UK
\\ 
$^2$ 
Physics Department, Boston University, Boston, MA 02215, USA
            }

\date{\today}

\begin{abstract}
We study a quantum phase transition between a phase which is
topologically ordered and one which is not. We focus on a spin model,
an extension of the toric code, for which we obtain the exact ground
state for all values of the coupling constant that takes the system
across the phase transition. We compute the entanglement and the
topological entropy of the system as a function of this coupling
constant, and show that the topological entropy remains constant all
the way up to the critical point, and jumps to zero beyond it. Despite
the jump in the topological entropy, the transition is second order as
detected via local observables.
\end{abstract}

\maketitle

\def\openone{\leavevmode\hbox{\small1\kern-4.2pt\normalsize1}}

\newcommand{\beq}{\begin{equation}}
\newcommand{\eeq}{\end{equation}}
\newcommand{\bea}{\begin{eqnarray}}
\newcommand{\eea}{\end{eqnarray}}
%
%

\section{\label{sec: intro}
Introduction
        } 
Some strongly correlated quantum many body systems display a type of
order which cannot be characterized by any local order
parameter. Instead, such order is topological in nature~\cite{topo
refs}, with the fractional quantum Hall systems being the primary
example so far: they are liquid states that exhibit exotic properties
such as a ground state (GS) degeneracy that cannot be lifted by any
local perturbations~\cite{Haldane1985,Wen1990} and fractionalized
degrees of freedom~\cite{Arovas1984}. Topologically ordered states are
also interesting in that their robustness against local perturbations
might be of use for decoherence-free quantum
computation~\cite{Kitaev2003}.

An example of an exactly solvable lattice spin model that is
topologically ordered was presented by Kitaev in 
Ref.~\onlinecite{Kitaev2003}, and
the system was argued to be robust against small perturbations that
tend to order the system \textit{\`{a} la} Landau-Ginzburg and take it 
away from its topological phase. 
The departure from the topologically ordered phase should occur through 
a quantum phase transition. Such quantum phase transition, however, 
cannot be entirely captured by ordinary methods based on
local Landau-Ginzburg order parameters, and new methods need to be devised 
in order to investigate the fate of topological order 
across the phase transition. 
These novel methods must be based on the fundamental properties 
of topologically ordered phases, such as the GS degeneracy in presence 
of a gap, and the presence of a non-vanishing topological entropy. 

Recent efforts to understand quantum phase transitions in topologically 
ordered states include a mean-field approach for these exotic 
states~\cite{Levin-WenMFT}, and analytical and numerical 
studies~\cite{Hamma2006,Trebst2006,Hamma2007} of the Kitaev model in the 
presence of a field. The numerical analysis presented in 
Refs.~\onlinecite{Trebst2006,Hamma2007} leads to the conclusion that 
topological order survives unchanged up to the second order phase 
transition at $\beta^{\ }_{c} = 0.32847(6)$ (in the notation of 
Eq.~(\ref{eq: Kitaev in field})), while the system is no longer 
topologically ordered for $\beta > \beta^{\ }_{c}$. (Here $\beta$ stands for 
the coupling constant that drives the $T=0$ quantum phase transition -- the
notation will become apparent shortly, and is chosen because of a
close relation to a classical model.)

In this paper, we investigate analytically a quantum phase transition
out of a topological phase. We show that the recently defined
topological entropy~\cite{Levin2006,Kitaev2006} works well as an 
``order parameter'' across the transition. We study the
transition using a model -- see Eq.~(\ref{eq: Kitaev field Ham}) --
that is shown to behave much like the Kitaev model in a magnetic field
for small values of the field. The advantage of this model is that the
ground state can be obtained exactly, from which we can then compute
the topological entropy explicitly, and show that it remains constant
in the topologically ordered phase 
($\beta < \beta^{\ }_{c} \simeq 0.4406868$), 
dropping abruptly to zero in the 
non-topologically-ordered phase 
($\beta > \beta^{\ }_{c}$), despite the continuous (second order) character 
of the transition. 

We find that in this model, even though one cannot identify a local
order parameter that vanishes in one phase and not in the other, one
can show that the (continuous) local magnetization has a singularity at
the critical point. In the model, we show that the magnetization
equals the energy $E_{\rm Ising}(\beta)$ of a $2D$ classical Ising model
with $N$ spins evaluated at an inverse (classical) temperature equal
to the value of the coupling constant $\beta$ that drives the system
through the $T=0$ phase transition:
\beq
m(\beta)=\frac{1}{N}\sum_i
\langle \hat{\sigma}^{\textrm{z}}_{i} \rangle
=\frac{1}{N} \,E_{\rm Ising}(\beta)
\;.
\eeq
{}From this relation, it becomes evident that the magnetization $m(\beta)$, 
although continuous and non-vanishing across the transition at $\beta_c$
(much as the energy of the classical Ising model across the
classical transition), has a singularity in its first derivative,
since 
\beq 
\frac{\partial m}{\partial \beta} 
= 
\frac{1}{N}\,\frac{\partial E_{\rm Ising}}{\partial \beta}
= 
-\beta^2\;\frac{1}{N}\,C_{\rm Ising}(\beta) 
\;, 
\eeq 
and the Ising model heat capacity $C_{\rm Ising}$ diverges
logarithmically at $\beta_c$. Hence, although there is no local order
parameter that can detect either the topological {\it or} the
non-topological phase in this system, one can expose the topological 
quantum phase transition through the singularity in the
derivative of a local quantity. 

This is contrasted, for example, with the case discussed in 
Sec.~\ref{sec: beyond 1-body potentials}, where a similar topological 
transition is accompanied by a simultaneous $\mathbb{Z}^{\ }_{2}$ symmetry 
breaking phase transition. In that case, the same transition is captured 
both by the non-local topological entropy, and by a local (Landau-Ginzburg) 
order parameter. 
%
%

\section{\label{sec: the model} 
The model
        } 
\begin{figure}[ht]
\vspace{0.2 cm}
\begin{center}
\includegraphics[width=0.7\columnwidth]{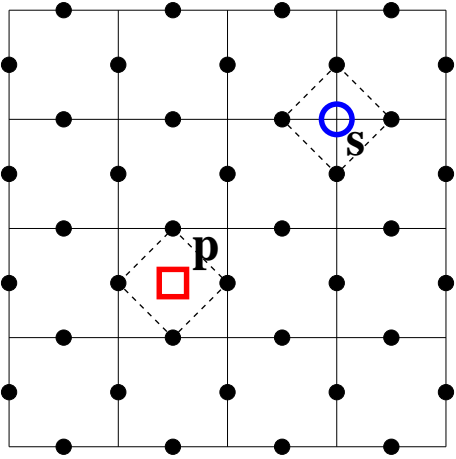}
\end{center}
\caption{
\label{fig: lattice operators}
(Color online) -- 
Examples of star and plaquette operators, centered at a lattice site $s$ 
(blue open circle) and at a dual lattice site $p$ (red open square), 
respectively. 
The solid black dots represent the spin-$1/2$ degrees of freedom 
living on the bonds of the lattice, and the dashed lines connect the spins 
involved in the definition of each of the above operators. 
}
\end{figure}

The model that we consider is a deformation of the Kitaev
model~\cite{Kitaev2003}, and it is defined on a square lattice with
spin-$1/2$ degrees of freedom living on the bonds, as shown in
Fig.~\ref{fig: lattice operators}. The pure Kitaev model is written in
terms of star and plaquette operators (see Fig.~\ref{fig: lattice
operators}). Star operators are defined as
\beq
A^{\ }_{s} 
= 
\prod^{\ }_{i \in \textrm{star}(s)} 
  \hat{\sigma}^{\textrm{x}}_{i} 
\equiv 
\prod^{\ }_{i \in s} 
  \hat{\sigma}^{\textrm{x}}_{i}, 
\eeq
where $i$ labels the four spins on the bonds departing from some
vertex $s$ of the square lattice. Plaquette operators are defined as
\beq
B^{\ }_{p} 
= 
  \prod^{\ }_{i \in \textrm{plaquette}(p)} \hat{\sigma}^{\textrm{z}}_{i} 
\equiv 
  \prod^{\ }_{i \in p} \hat{\sigma}^{\textrm{z}}_{i},
\eeq
where $i$ labels the four spins on the bonds around some plaquette $p$
of the square lattice.

The Hamiltonian we consider in this paper is
\bea
H 
&=& 
- 
\lambda^{\ }_{0} 
  \sum^{\ }_{p} 
    B^{\ }_{p} 
- 
\lambda^{\ }_{1} 
\sum^{\ }_{s} 
    A^{\ }_{s} 
+ 
\lambda^{\ }_{1}
\sum^{\ }_{s} 
  e^{- \beta \sum^{\ }_{i \in s} \hat{\sigma}^{\textrm{z}}_{i}}_{\ } 
\nonumber\\ 
&=& 
H^{\ }_{\textrm{Kitaev}} 
+ 
\lambda^{\ }_{1}
\sum^{\ }_{s} 
  e^{- \beta \sum^{\ }_{i \in s} \hat{\sigma}^{\textrm{z}}_{i} }_{\ },
\label{eq: Kitaev field Ham}
\eea
where $\lambda_{0,1}>0$ and $\beta$ is a parameter that we use to tune the 
system across a topological quantum phase transition. 
Notice that for $\beta=0$ the
Hamiltonian 
Eq.~(\ref{eq: Kitaev field Ham}) 
is simply the Kitaev Hamiltonian $H^{\ }_{\textrm{Kitaev}}$ in 
Ref.\onlinecite{Kitaev2003}, up to a trivial overall constant shift 
of the energy. 

The exact ground state wavefunction of this Hamiltonian can be obtained 
by deconstructing $H$ into two pieces,
$H=\lambda_0 H_0 + \lambda_1 H_1$, as follows. 

Take $G$ to be the (Abelian) group of all spin flip
operations obtained as products of star type operators.
Notice that $g^{2}_{\ } = \openone$
for any element $g$ of the group $G$. By acting with elements of $G$ on a
given reference configuration $ \bigotimes^{\ }_{i} \vert
\sigma^{\textrm{z}}_{i} \rangle $ one generates a manifold of states,
which however does not encompass the whole basis. For example, the
action of a star operator $A^{\ }_{s}$ cannot change the sign of the
product of $\sigma^{\textrm{z}}_{\ }$'s around any square plaquette in
the lattice~\cite{Kitaev2003,Hamma2005} (see 
Fig.~\ref{fig: lattice operators}). Therefore, there is a
non-trivial (and non-unique) minimal set $\{ \vert \Psi^{\ }_{\alpha}
\rangle \}$ of reference configurations that generates the full
$\sigma^{\textrm{z}}_{\ }$-basis under the action of the group $G$.
(In particular, one of the elements in this set is the reference
configuration $\vert 0 \rangle$ that is fully magnetized in the
$z$-direction, say $\sigma^{\textrm{z}}_{i} = 1$, $\forall i$.)

Consider then the family of Hamiltonians 
\beq
H^{\ }_{1}(\beta) 
= 
\sum^{\ }_{s} 
  \left[
    e^{- \beta \sum^{\ }_{i \in s} \hat{\sigma}^{\textrm{z}}_{i} }_{\ } 
    - 
    \prod^{\ }_{i \in s} \hat{\sigma}^{\textrm{x}}_{i} 
  \right], 
\label{eq: H1}
\eeq
for some real-valued parameter $\beta$. The ground state  
of any such Hamiltonian can be obtained exactly and it can 
be written in the form 
\bea
\vert GS^{\ }_{1} \rangle 
&=& 
\sum^{\ }_{\alpha} 
  \psi^{\ }_{\alpha} \sum^{\ }_{g \in G} 
    \frac{e^{\beta \sum^{\ }_{i} \sigma^{\textrm{z}}_{i}(g,\alpha) / 2}}
         {\sqrt{Z^{\ }_{\alpha}}} 
      g \, \vert \Psi^{\ }_{\alpha} \rangle, 
\\ 
Z^{\ }_{\alpha} 
&=& 
\sum^{\ }_{g \in G} 
  e^{\beta \sum^{\ }_{i} \sigma^{\textrm{z}}_{i}(g,\alpha)} 
\label{eq: H1 GS}
\eea
where $\alpha$ labels the different block-diagonal sectors corresponding to 
the states in the minimal set $\{ \vert \Psi^{\ }_{\alpha} \rangle \}$; 
$\sigma^{\textrm{z}}_{i}(g,\alpha)$ is the $z$-component of the spin at site 
$i$ in state $g \, \vert \Psi^{\ }_{\alpha} \rangle$; and 
the coefficients $\psi^{\ }_{\alpha}$ can be chosen at will, 
subject to the normalization condition 
$\sum^{\ }_{\alpha} \vert \psi_{\alpha} \vert^{2} = 1$. 
Although the choice of minimal set is non-unique, one can show that 
Eq.~(\ref{eq: H1 GS}) is independent of such choice, modulo an irrelevant 
permutation of the $\alpha$ indices. 
Within each block-diagonal sector, the 
GS of Eq.~(\ref{eq: H1}) is unique.  Instead of proving directly
that~(\ref{eq: H1 GS}) is the GS of~(\ref{eq: H1}), it is more
convenient to notice that the family of Hamiltonians in Eq.~(\ref{eq:
H1}) is a particular choice of Stochastic Matrix Form decompositions
of quantum Hamiltonians that exhibit precisely Eq.~(\ref{eq: H1 GS})
as their GS.~\cite{Castelnovo2005} One can verify this by showing that
each of the operators
\beq
Q_s=e^{- \beta \sum^{\ }_{i \in s}
\hat{\sigma}^{\textrm{z}}_{i} }_{\ } - \prod^{\ }_{i \in s}
\hat{\sigma}^{\textrm{x}}_{i} 
\eeq
between square brackets in Eq.~(\ref{eq: H1}) annihilates the inner sum in
$\vert GS^{\ }_{1} \rangle$, independently of the index $\alpha$. 
That the GS energy is zero follows because 
\beq
{Q_s}^2=
2\cosh\left(\beta \sum^{\ }_{i\in s} \hat{\sigma}^{\textrm{z}}_{i}\right)
\;
Q_s
\eeq
and
\beq
\left[Q_s,
\cosh\left(\beta \sum^{\ }_{i\in s} \hat{\sigma}^{\textrm{z}}_{i}\right)
\right]=0
\;,
\eeq
from which it can be shown that the expectation value of $Q_s$ with
respect to any state is always greater than or equal to zero.

Let us consider now the remaining part of the Hamiltonian, 
\beq
H^{\ }_{0} 
= 
- 
\sum^{\ }_{p} 
  \prod^{\ }_{i \in \textrm{plaquette}(p)} \hat{\sigma}^{\textrm{z}}_{i} 
\equiv 
- 
\sum^{\ }_{p} 
  \prod^{\ }_{i \in p} \hat{\sigma}^{\textrm{z}}_{i}. 
\label{eq: H0}
\eeq
Recall that any star operator $A^{\ }_{s}$, and therefore any element of 
the group $G$, preserves the product 
$\prod^{\ }_{i \in p} \hat{\sigma}^{\textrm{z}}_{i}$ on every plaquette of 
the lattice. 
The GS wavefunction of Hamiltonian~(\ref{eq: H0}) can then be written as 
\beq
\vert GS^{\ }_{0} \rangle 
= 
{\sum^{\ }_{\alpha}}^{\prime} \sum^{\ }_{g \in G} 
  \phi^{\ }_{g,\alpha} \, 
    g \, \vert \Psi^{\ }_{\alpha} \rangle, 
\label{eq: H0 GS}
\eeq
for any choice of the coefficients $\phi^{\ }_{g,\alpha}$ 
(${\sum^{\ }_{\alpha}}^{\prime} \sum^{\ }_{g \in G} 
\vert \phi^{\ }_{g,\alpha} \vert^{2} = 1$). 
Here the primed sum over $\alpha$ is restricted to the (four) block-diagonal 
sectors that satisfy 
$\prod^{\ }_{i \in p} \hat{\sigma}^{\textrm{z}}_{i} = +1$ 
for all plaquettes $p$ in the lattice, and it must be carried out separately 
because no operation in $G$ allows to change sector.~\cite{Kitaev2003} 

As a result, any linear combination with positive weights 
$\lambda^{\ }_{0}$ and $\lambda^{\ }_{1}$, 
\bea
H &=& \lambda^{\ }_{0} H^{\ }_{0} + \lambda^{\ }_{1} H^{\ }_{1}
\nonumber\\
&=&
H^{\ }_{\textrm{Kitaev}} 
+ 
\lambda^{\ }_{1}
\sum^{\ }_{s} 
  e^{- \beta \sum^{\ }_{i \in s} \hat{\sigma}^{\textrm{z}}_{i} }_{\ },
\label{eq: Kitaev field Ham1}
\eea
and therefore our  Hamiltonian in Eq.~(\ref{eq: Kitaev field Ham}),
has the GS given by 
\beq
\vert GS \rangle 
= 
{\sum^{\ }_{\alpha}}^{\prime} \psi^{\ }_{\alpha} 
  \sum^{\ }_{g \in G} 
    \frac{e^{\beta \sum^{\ }_{i} \sigma^{\textrm{z}}_{i}(g,\alpha) / 2}}
         {\sqrt{Z^{\ }_{\alpha}}} 
      g \, \vert \Psi^{\ }_{\alpha} \rangle. 
\label{eq: Kitaev field GS}
\eeq
Notice that one of the topological sectors that satisfy $\prod^{\ }_{i
\in p} \hat{\sigma}^{\textrm{z}}_{i}=+1$, $\forall\,p$ is the one containing 
the fully magnetized configuration in the $z$-direction ($\vert 0
\rangle$).

In particular for
$\vert \beta \vert \ll 1$
\bea
e^{- \beta \sum^{\ }_{i \in s} \hat{\sigma}^{\textrm{z}}_{i}}_{\ } 
&\simeq& 
1 - \beta \sum^{\ }_{i \in s} \hat{\sigma}^{\textrm{z}}_{i} 
\\ 
\lambda^{\ }_{1}
\sum^{\ }_{s} 
  e^{- \beta \sum^{\ }_{i \in s} \hat{\sigma}^{\textrm{z}}_{i} }_{\ } 
&\simeq& 
\textrm{const} 
-
2 \beta \lambda^{\ }_{1} 
  \sum^{\ }_{i} \hat{\sigma}^{\textrm{z}}_{i} . 
\eea
Therefore, in the limit of small $\beta$ (in absolute value) the Hamiltonian 
in Eqs.~(\ref{eq: Kitaev field Ham},\ref{eq: Kitaev field Ham1}) 
is equivalent to the Kitaev model in 
presence of a magnetic field proportional to $\beta \lambda^{\ }_{1}$, 
\beq
H 
= 
- 
\lambda^{\ }_{0} 
  \sum^{\ }_{p} 
    \prod^{\ }_{i \in p} \hat{\sigma}^{\textrm{z}}_{i} 
- 
\lambda^{\ }_{1} 
\sum^{\ }_{s} 
  \prod^{\ }_{i \in s} \hat{\sigma}^{\textrm{x}}_{i} 
-
2 \beta \lambda^{\ }_{1} 
  \sum^{\ }_{i} \hat{\sigma}^{\textrm{z}}_{i} . 
\label{eq: Kitaev in field}
\eeq
For larger values of $\beta$, the many-body terms in 
Eq.~(\ref{eq: SMF field generated couplings}) are no longer negligible and 
the equivalence is lost, although the form of the 
GS~(\ref{eq: Kitaev field GS}) suggests that the system gets deeper and 
deeper into the magnetized phase -- as one would expect upon increasing the 
strength of the magnetic field in the Kitaev model. 
As we discuss in Section~\ref{sec: 1-body potentials}, our model undergoes 
a second-order phase transition at 
$\beta^{\ }_{c} = (1/2) \ln (\sqrt{2} + 1) \simeq 0.4406868$, 
where it displays a dimensionality reduction that places the transition in 
a different universality class than the one studied in 
Refs.~\onlinecite{Hamma2006,Trebst2006,Hamma2007}. 

One can use the decomposition 
\bea
e^{- \beta \sum^{\ }_{i \in s} \hat{\sigma}^{\textrm{z}}_{i}}_{\ } 
&=& 
\prod^{\ }_{i \in s} 
  \left[\vphantom{\sum} 
    \cosh(\beta) 
    - 
    \hat{\sigma}^{\textrm{z}}_{i} \sinh(\beta)
  \right] 
\nonumber\\ 
&=& 
\cosh^{4}_{\ }(\beta) 
\nonumber\\ 
&-&
\cosh^{3}_{\ }(\beta) \sinh(\beta) 
  \sum^{\ }_{i \in s} \hat{\sigma}^{\textrm{z}}_{i} 
\nonumber\\ 
&+&
\cosh^{2}_{\ }(\beta) \sinh^{2}_{\ }(\beta) 
  \sum^{\ }_{i \neq j \in s} 
    \hat{\sigma}^{\textrm{z}}_{i} \hat{\sigma}^{\textrm{z}}_{j} 
\nonumber\\ 
&-& 
\cosh(\beta) \sinh^{3}_{\ }(\beta) 
  \sum^{\ }_{i \neq j \neq k \in s} 
    \hat{\sigma}^{\textrm{z}}_{i} 
    \hat{\sigma}^{\textrm{z}}_{j} 
    \hat{\sigma}^{\textrm{z}}_{k} 
\nonumber\\ 
&+& 
\sinh^{4}_{\ }(\beta) 
  \prod^{\ }_{i \in s} \hat{\sigma}^{\textrm{z}}_{i}, 
\label{eq: SMF field generated couplings}
\eea
to estimate the limit of validity of Eq.~(\ref{eq: Kitaev in field}) 
to be given by the condition 
\beq
\left\vert
  \frac{\cosh^{3}_{\ }(\beta) \sinh(\beta)}
       {\cosh^{2}_{\ }(\beta) \sinh^{2}_{\ }(\beta)}
\right\vert
= 
\left\vert
  \frac{\cosh(\beta)}{\sinh(\beta)} 
\right\vert
\gtrsim 
2. 
\eeq
This corresponds to a ratio between the coupling to the magnetic field 
and the coupling to the cooperative transverse field 
($
\sum^{\ }_{s} 
  \prod^{\ }_{i \in s} \hat{\sigma}^{\textrm{x}}_{i}
$) 
\beq
\left\vert
\frac{2 \lambda^{\ }_{1} \cosh^{3}_{\ }(\beta) \sinh(\beta)}
     {\lambda^{\ }_{1}}
\right\vert
\lesssim 
\frac{16}{9}
\simeq
1.78. 
\label{eq: field to hopping ratio}
\eeq
The detailed numerical analysis presented in 
Refs.~\onlinecite{Trebst2006,Hamma2007} 
lead the authors to conclude that topological order survives up to the 
second order phase transition at finite $\beta^{\ }_{c}$ 
(in the notation of Eq.~(\ref{eq: Kitaev in field})), 
while the system is no longer topologically ordered for 
$\beta > \beta^{\ }_{c}$. 
In the following, we investigate this phase and the relative phase 
transition using the exact ground state of our 
model~(\ref{eq: Kitaev field Ham}) to compute the topological 
entropy~\cite{Levin2006,Kitaev2006} across the transition. 
Using an exact derivation from the microscopic degrees of freedom, we show 
that the topological entropy is able to detect a transition from a 
topologically ordered phase ($\beta < \beta^{\ }_{c}$) 
to a non-topologically-ordered phase 
($\beta > \beta^{\ }_{c}$). 
Indeed, it remains constant at its known 
$\beta \to 0$ value up to the transition and drops abruptly to zero 
afterwards, despite the continuous character of the transition. 
%
%

\section{\label{sec: topo entropy of factorizable wavefunctions}
The topological entropy of factorizable (local) wavefunctions
        } 
\begin{figure}[ht]
\vspace{0.2 cm}
\includegraphics[width=0.98\columnwidth]{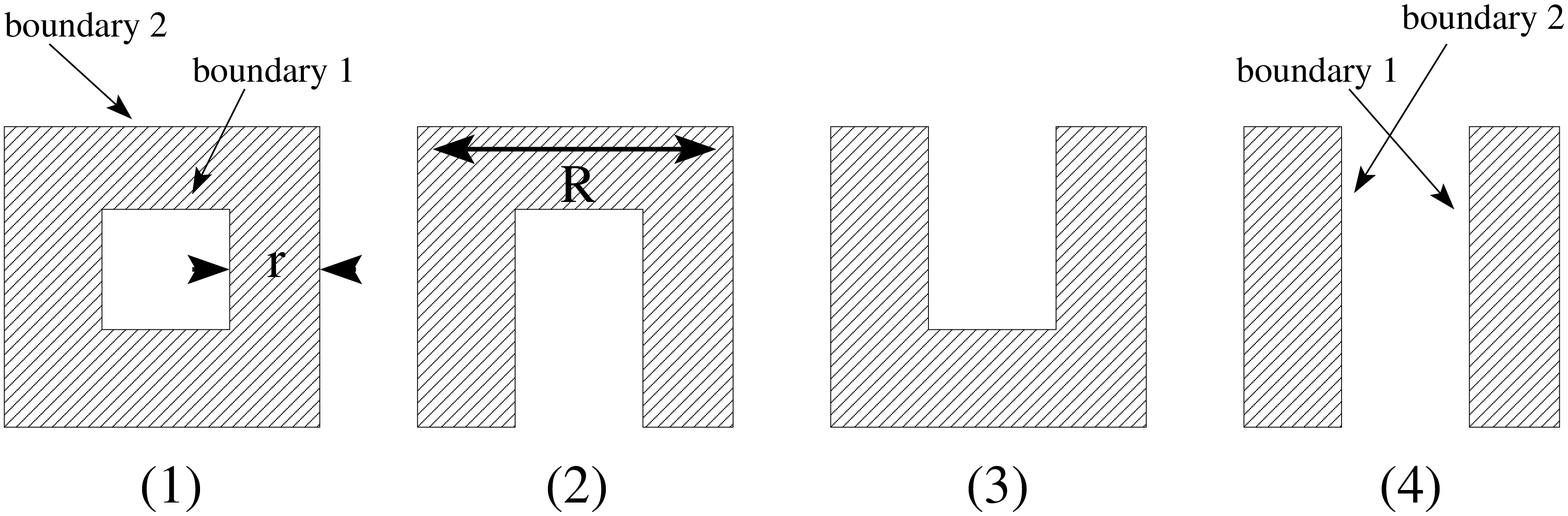}
\caption{
\label{fig: topological partitions}
Illustration of the four bipartitions used to compute the topological 
entropy in Ref.~\onlinecite{Levin2006}. 
}
\end{figure}

Using the definition in Refs.~\onlinecite{Levin2006,Kitaev2006}, the 
topological entropy can be obtained as a linear combination of Von Neumann 
entanglement entropies $S^{\ }_{\textrm{VN}}$ of different bipartitions of 
the system into subsystems $A$ and $B$: 
\beq
S^{A}_{\textrm{VN}} 
\equiv 
-\textrm{Tr} \left[ \rho^{\ }_{A} \log^{\ }_{2} \rho^{\ }_{A} \right] 
= 
S^{B}_{\textrm{VN}}, 
\label{eq: S_VN}
\eeq
where $\rho^{\ }_{A} = \textrm{Tr}^{\ }_{B} (\rho)$ is the reduced density 
matrix obtained from the full density matrix $\rho$ by tracing out the 
degrees of freedom of subsystem $B$, and the last equality holds whenever 
the full density matrix $\rho$ is a pure-state density matrix. 
The different bipartitions are aimed at removing all 
the extensive (boundary) contributions to uncover the sole topological 
contribution. 
A particular choice of the four bipartitions~\cite{Levin2006} is illustrated 
in Fig.~\ref{fig: topological partitions}, 
and the topological entropy is then defined
as:
\beq
S^{\ }_{\textrm{topo}} 
= 
\lim^{\ }_{r, R \to \infty} 
\left( 
  - S^{(1A)}_{\textrm{VN}} 
  + S^{(2A)}_{\textrm{VN}} 
  + S^{(3A)}_{\textrm{VN}} 
  - S^{(4A)}_{\textrm{VN}} 
\right).
\label{eq: topo entropy}
\eeq

In order to compute the topological entropy as a function of the parameter 
$\beta$, let us first notice that 
the Hamiltonian~(\ref{eq: Kitaev field Ham1}) constructed above, 
with GS given by~(\ref{eq: Kitaev field GS}), belongs to a class of 
Hamiltonians whose GS wavefunctions 
$
\vert \Psi \rangle 
= 
\vert Z \vert^{-1/2}_{\ } 
\sum^{\ }_{g \in G} e^{-\beta E^{\ }_{g} / 2}_{\ } \,g \vert 0 \rangle
$ 
have non-negative, factorizable amplitudes, i.e., 
$E^{\ }_{g} = E^{A}_{g^{\ }_{A}} + E^{B}_{g^{\ }_{B}}$, 
with $g = g^{\ }_{A} \otimes g^{\ }_{B}$ for all bipartitions $(A,B)$. 
For this type of Hamiltonians, one can compute the entanglement entropy 
as follows. 

Consider a given bipartition $(A,B)$ of the system. 
The reduced density matrix 
$
\rho^{\ }_{A} 
= 
\textrm{Tr}^{\ }_{B} \left( \rho \right) 
$, 
obtained by tracing over all degrees of freedom 
in $B$, is given by~\cite{Hamma2005} 
\bea
\rho^{\ }_{A} 
&=& 
\frac{1}{Z} 
  \sum^{\ }_{g,\tilde{g} \in G} 
    e^{-\beta (E^{\ }_{g} + E^{\ }_{g \tilde{g}})/2} 
\nonumber \\ 
&\times& 
    \langle 0^{\ }_{B} \vert g^{\ }_{B} 
      \tilde{g}^{\ }_{B} g^{\ }_{B} \vert 0^{\ }_{B} \rangle 
    \;\; 
    g^{\ }_{A} \vert 0^{\ }_{A} \rangle 
      \langle 0^{\ }_{A} \vert g^{\ }_{A} \tilde{g}^{\ }_{A} 
\nonumber \\ 
&=& 
\frac{1}{Z} 
  \sum^{\ }_{g \in G,\, g^{\prime}_{\ } \in G^{\ }_{A}} 
    e^{-\beta E^{B}_{g^{\ }_{B}}} 
    e^{-\beta (E^{A}_{g^{\ }_{A}} + E^{A}_{g^{\ }_{A} g^{\prime}_{A}})/2} 
\nonumber \\ 
&\times&
    g^{\ }_{A} 
      \vert 0^{\ }_{A} \rangle \langle 0^{\ }_{A} \vert 
        g^{\ }_{A} g^{\prime}_{A}. 
\nonumber\\ 
\label{eq: rho_A}
\eea
where $g=g^{\ }_{A} \otimes g^{\ }_{B}$, 
$\vert 0 \rangle = \vert 0^{\ }_{A} \rangle \otimes \vert 0^{\ }_{B} \rangle$ 
and $G^{\ }_{A} \subset G$ ($G^{\ }_{B} \subset G$) is the subgroup of 
transformations acting solely on $A$ ($B$) and leaving $B$ ($A$) invariant: 
\bea
G^{\ }_{A} 
&=& 
\{ g \in G \;\vert\; g^{\ }_{B} = \openone^{\ }_{B} \}
\nonumber \\
G^{\ }_{B} 
&=& 
\{ g \in G \;\vert\; g^{\ }_{A} = \openone^{\ }_{A} \}
\nonumber 
\eea
Notice that we used the group property to rewrite a generic element of $G$ 
as $g \tilde{g}$, $\exists ! \, \tilde{g} \in G$, as well as the additive 
property of $E^{\ }_{g}$. 

We can then compute the trace of the $n$-th power of the reduced density 
matrix 
$
\textrm{Tr} 
  \left[ 
    \left( 
      \rho^{\ }_{A} 
    \right)^{n}_{\ } 
  \right]
$ 
and use the identity 
\beq
- 
\lim^{\ }_{n \to 1} 
  \frac{\partial}{\partial n}
    \textrm{Tr} 
      \left[ 
        \left( 
          \rho^{\ }_{A} 
        \right)^{n}_{\ } 
      \right]
= 
- 
\textrm{Tr} 
  \left[ 
    \rho^{\ }_{A} \ln \rho^{\ }_{A}
  \right] 
\label{eq: trace identity}
\eeq
to obtain the Von Neumann entropy 
$
S^{(A)}_{\textrm{VN}} 
= 
- 
\textrm{Tr} 
  \left[ 
    \rho^{\ }_{A} \log^{\ }_{2} \rho^{\ }_{A}
  \right]
$: 
\bea
\textrm{Tr} 
  \left[ 
    \rho^{n}_{A} 
  \right] 
&=& 
\frac{1}{Z^{n}_{\ }} 
  \mathop{\sum^{\ }_{g^{\ }_{1}, \ldots, g^{\ }_{n} 
          \in G}}_{g^{\prime}_{1}, \ldots, g^{\prime}_{n} 
	  \in G^{\ }_{A}} 
    e^{-\beta \sum^{n}_{i=1} E^{B}_{g^{\ }_{i,B}}} 
\nonumber \\ 
&& 
\qquad\qquad\quad\; 
\times \; 
    e^{-\beta \sum^{n}_{i=1} 
       (E^{A}_{g^{\ }_{i,A}} + E^{A}_{g^{\ }_{i,A} g^{\prime}_{i,A}})/2} 
\nonumber \\ 
&\times& 
    \langle 0^{\ }_{A} \vert g^{\ }_{1,A} g^{\prime}_{1,A} 
      g^{\ }_{2,A} \vert 0^{\ }_{A} \rangle 
    \langle 0^{\ }_{A} \vert g^{\ }_{2,A} g^{\prime}_{2,A} 
      g^{\ }_{3,A} \vert 0^{\ }_{A} \rangle 
    \ldots 
\nonumber \\ 
&\times&
  \ldots 
    \langle 0^{\ }_{A} \vert g^{\ }_{n,A} g^{\prime}_{n,A} 
      g^{\ }_{1,A} \vert 0^{\ }_{A} \rangle 
\nonumber \\ 
&=& 
\frac{1}{Z^{n}_{\ }} 
  \mathop{\sum^{\ }_{g^{\ }_{1}, \ldots, g^{\ }_{n} 
          \in G}}_{g^{\prime}_{1}, \ldots, g^{\prime}_{n} 
	  \in G^{\ }_{A}} 
    e^{-\beta \sum^{n}_{i=1} E^{\ }_{g^{\ }_{i}}} 
\nonumber \\ 
&\times& 
    \langle 0^{\ }_{A} \vert g^{\ }_{1,A} g^{\prime}_{1,A} 
      g^{\ }_{2,A} \vert 0^{\ }_{A} \rangle 
    \langle 0^{\ }_{A} \vert g^{\ }_{2,A} g^{\prime}_{2,A} 
      g^{\ }_{3,A} \vert 0^{\ }_{A} \rangle 
    \ldots 
\nonumber \\ 
&\times&
  \ldots 
    \langle 0^{\ }_{A} \vert g^{\ }_{n,A} g^{\prime}_{n,A} 
      g^{\ }_{1,A} \vert 0^{\ }_{A} \rangle 
, 
\nonumber \\ 
\label{eq: trace rho n quantum field 1}
\eea
where we used the fact that the inner products in 
Eq.~(\ref{eq: trace rho n quantum field 1}) impose 
\beq
g^{\ }_{i+1,A} = g^{\ }_{i,A} g^{\prime}_{i,A} 
, 
\label{eq: inner prod conditions}
\eeq
for $i=1,\ldots,n$, with the identification $n+1 \equiv 1$, 
and therefore 
\beq
E^{B}_{g^{\ }_{i,B}} 
+ 
\frac{1}{2}
(E^{A}_{g^{\ }_{i,A}} + E^{A}_{g^{\ }_{i-1,A} g^{\prime}_{i-1,A}}) 
= 
E^{\ }_{g^{\ }_{i}} 
. 
\eeq
The condition in Eq.~(\ref{eq: inner prod conditions}) can be satisfied if 
and only if 
\beq
g^{\prime}_{i} 
= 
\left( g^{\ }_{i+1,A} g^{\ }_{i,A} \right) 
\otimes 
\openone^{\ }_{B} 
\in G^{\ }_{A} 
. 
\eeq
Thus, the summation over all $g^{\prime}_{i}$ of the inner products in 
Eq.~(\ref{eq: trace rho n quantum field 1}) yields a constraint over the 
allowed values of $g^{\ }_{1}, \ldots, g^{\ }_{n} \in G$: 
\beq
\begin{array}{c}
\left( g^{\ }_{i+1,A} g^{\ }_{i,A} \right) 
\otimes 
\openone^{\ }_{B} 
\in G^{\ }_{A}, 
\qquad 
\forall \, i=1,\ldots,n 
\;\; 
(n+1 \equiv 1) 
\\
\Updownarrow 
\\ 
\left( g^{\ }_{i,A} g^{\ }_{j,A} \right) 
\otimes 
\openone^{\ }_{B} 
\in G^{\ }_{A}, 
\qquad 
\forall \, i,j=1,\ldots,n 
\\
\Updownarrow 
\\ 
g^{\ }_{i} g^{\ }_{j} 
\in G^{\ }_{A}G^{\ }_{B}, 
\qquad 
\forall \, i,j=1,\ldots,n 
. 
\end{array} 
\eeq
In particular, the last line can be recast as 
\beq
g^{\ }_{i} = h^{\ }_{i} \, g^{\ }_{1} \, k^{\ }_{i} 
\;\;\;\; 
\exists ! \, h^{\ }_{i} \in G^{\ }_{A}, \; k^{\ }_{i} \in G^{\ }_{B}, 
\;\; 
\forall \, i=2,\ldots,n 
. 
\label{eq: boundary constraint}
\eeq
The physical meaning of these conditions will become clear in the next 
section, for the specific case of the system considered in this paper, 
although the form of Eq.~(\ref{eq: boundary constraint}) already suggests 
that they require all the elements $g^{\ }_{i}$ to agree at the boundary of 
the bipartition $(A,B)$. 

We can finally use Eq.~(\ref{eq: boundary constraint}) to simplify 
Eq.~(\ref{eq: trace rho n quantum field 1}) 
\bea
\textrm{Tr} 
  \left[ 
    \rho^{n}_{A} 
  \right] 
&=& 
\frac{1}{Z^{n}_{\ }} 
  \sum^{\ }_{g \in G} e^{-\beta E^{\ }_{g}} 
    \mathop{\sum^{\ }_{h^{\ }_{2}, \ldots, h^{\ }_{n} 
        \in G^{\ }_{A}}}^{\ }_{k^{\ }_{2}, \ldots, k^{\ }_{n} 
        \in G^{\ }_{B}} 
    e^{-\beta \sum^{n}_{i=2} E^{\ }_{h^{\ }_{i} g k^{\ }_{i}}} 
\nonumber \\ 
&=& 
\frac{1}{Z} 
  \sum^{\ }_{g \in G} e^{-\beta E^{\ }_{g}} 
   \left( 
    \frac{\sum^{\ }_{h \in G^{\ }_{A},\, k \in G^{\ }_{B}} 
            e^{-\beta E^{\ }_{h g k}}}
	 {Z} 
     \right)^{n-1}_{\ } 
\label{eq: trace rho n quantum field 3}
\eea
and obtain, via Eq.~(\ref{eq: trace identity}), 
\bea
S^{(A)}_{\textrm{VN}} 
&=& 
- \frac{1}{Z}
  \sum^{\ }_{g \in G} e^{-\beta E^{\ }_{g}} 
    \log^{\ }_{2} 
      \left[
        \frac{\sum^{\ }_{h \in G^{\ }_{A},\, k \in G^{\ }_{B}} 
              e^{-\beta E^{\ }_{h g k}}}
	     {Z} 
      \right] 
\nonumber\\ 
&=&
\left\langle 
  \log^{\ }_{2} 
    \left[
      \frac{\sum^{\ }_{h \in G^{\ }_{A},\, k \in G^{\ }_{B}} 
            e^{-\beta E^{\ }_{h g k}}}
	   {Z} 
    \right] 
\right\rangle
\nonumber\\ 
&=&
- 
\langle \log^{\ }_{2} \tilde{Z}^{\ }_{g} \rangle 
+ 
\log^{\ }_{2} Z 
= 
\beta 
  \left( 
  \langle \tilde{F}^{\ }_{g} \rangle 
  - 
  F 
  \right), 
\label{eq: 1-body von Neumann entropy}
\eea
where $\tilde{F}^{\ }_{g}$ is the \textit{partial} free energy given by 
all the configurations that can be obtained from $g$ via products of spin 
flip operators that act solely on subsystem $A$ or subsystem $B$ 
(i.e., having the same `boundary' as $g$), 
and $\langle \ldots \rangle$ denotes the ensemble average over $g \in G$ 
with weight $e^{-\beta E^{\ }_{g}}$. 
Notice that our result in Eq.~(\ref{eq: 1-body von Neumann entropy}) is the 
lattice equivalent of the Von Neumann entropy obtained by Fradkin and Moore 
in Ref.~\onlinecite{Fradkin2006} for continuous systems.

Alternatively, Eq.~(\ref{eq: 1-body von Neumann entropy}) can be interpreted 
as the entropy of mixing (or configurational entropy) of the allowed 
bipartition boundaries in $G$. 
This can be made more transparent by introducing the quotient group 
$Q = G / (G^{\ }_{A} G^{\ }_{B})$, and by rewriting 
Eq.~(\ref{eq: 1-body von Neumann entropy}) as 
\bea
S^{(A)}_{\textrm{VN}} 
&=& 
- \sum^{\ }_{q \in Q} 
    \mathop{\sum^{\ }_{h \in G^{\ }_{A}}}^{\ }_{k \in G^{\ }_{B}} 
      \frac{e^{-\beta E^{\ }_{hqk}}}{Z} 
\nonumber \\ 
&& 
\qquad
\times 
    \log^{\ }_{2} 
      \left[
        \frac{\sum^{\ }_{\tilde{h} \in G^{\ }_{A},\, \tilde{k} \in G^{\ }_{B}} 
              e^{-\beta E^{\ }_{\tilde{h} (hqk) \tilde{k}}}}
	     {Z} 
      \right] 
\nonumber\\ 
&=& 
- \sum^{\ }_{q \in Q} 
    P^{\ }_{q} \: 
    \log^{\ }_{2} 
      P^{\ }_{q} 
, 
\eea
where we used the fact that the term in square brackets is independent of 
$h$ and $k$, and where we introduced the notation 
\beq
P^{\ }_{q} 
= 
\frac{\sum^{\ }_{h \in G^{\ }_{A},\, k \in G^{\ }_{B}} 
      e^{-\beta E^{\ }_{h q k}}}
     {Z} 
\eeq
for the probability of boundary $q$ to appear in $G$, for a given inverse 
temperature $\beta$ and energy $E^{\ }_{g}$. 

In order to proceed further, let us focus for simplicity on the specific 
GS of our system~(\ref{eq: Kitaev field GS}). 
The generic case of a wavefunction with factorizable amplitudes can be 
inferred with minor modifications. 
%
%

\subsection{\label{sec: 1-body potentials}
The case of $1$-body potentials
           } 
All of the above results apply straightforwardly to the GS in 
Eq.~(\ref{eq: Kitaev field GS}). Notice that 
(i) the topological entropy in each block-diagonal sector of the pure Kitaev 
model is the same,~\cite{Hamma2005} and
(ii) it is reasonable to make the working assumption that 
the relevant sector for the transition to the fully magnetized 
state $\vert 0 \rangle$ is the one that contains this state, and that is 
therefore obtained upon applying the group $G$ to $\vert 0 \rangle$. 
For the purpose of computing the topological entropy, one can thus replace 
Eq.~(\ref{eq: Kitaev field GS}) by 
\beq
\vert GS \rangle 
= 
\frac{1}{\sqrt{Z}} 
  \sum^{\ }_{g \in G} 
    e^{\beta \sum^{\ }_{i} \sigma^{\textrm{z}}_{i}(g) / 2}
      g \, \vert 0 \rangle. 
\eeq
and obtain 
\bea
S^{(A)}_{\textrm{VN}} 
&=& 
- \frac{1}{Z}
  \sum^{\ }_{g \in G} e^{\beta \sum^{\ }_{i} \sigma^{\textrm{z}}_{i}(g)} 
\nonumber \\ 
&\times&
    \log^{\ }_{2} 
      \left[
        \frac{\sum^{\ }_{h \in G^{\ }_{A},\, k \in G^{\ }_{B}} 
              e^{\beta \sum^{\ }_{i} \sigma^{\textrm{z}}_{i}(h g k)}}
	     {Z} 
      \right] 
\label{eq: von Neumann field entropy}
\eea
where 
$
Z 
= 
\sum^{\ }_{g \in G} e^{\beta \sum^{\ }_{i} \sigma^{\textrm{z}}_{i}(g)}
$. 

In order to simplify Eq.~(\ref{eq: von Neumann field entropy}) with the 
purpose of computing the topological entropy of the 
system~(\ref{eq: topo entropy}), it is convenient to do the following change 
of variables. Recall that a generic configuration $g\vert0\rangle$ is 
uniquely specified by the set of star operators acting on the reference 
configuration $\vert0\rangle$, which we chose to be the ferromagnetic state 
with all the $\sigma$ spins pointing up, modulo the action of the product of 
all the star operators (which is equal to the identity). 
Thus, there is a $1$-to-$2$ mapping between $G = \{ g \}$ and the 
configuration space $\Theta = \{ \bftheta \}$ of an Ising model with 
degrees of freedom $\theta^{\ }_{s}$ living on the sites $s$ of the square 
lattice, where for example $\theta^{\ }_{s} = -1$ ($+1$) means that the 
corresponding star operator is (not) acting in the associated $g$. 
Since each $\sigma$ spin can be flipped only by its two neighboring 
$\theta$ spins, then 
$
\sigma^{\ }_{i} 
\equiv 
\theta^{\ }_{s}\theta^{\ }_{s^{\prime}_{\ }}
$, 
where $i$ labels the bond between the two neighboring sites 
$\langle s,s^{\prime}_{\ } \rangle$, and 
\beq
\sum^{\ }_{g \in G} 
  e^{\beta \sum^{\ }_{i} \sigma^{\textrm{z}}_{i}(g)} 
\equiv
\frac{1}{2} \sum^{\ }_{\bftheta \in \Theta} 
  e^{\beta \sum^{\ }_{\langle s,s^{\prime}_{\ } \rangle} 
     \theta^{\ }_{s} \theta^{\ }_{s^{\prime}_{\ }}}. 
\eeq

Notice that, using the above mapping, the GS wavefunction of our model, 
Eq.~(\ref{eq: Kitaev field GS}), can be rewritten as 
\beq
\vert GS \rangle 
= 
  \sum^{\ }_{\bftheta \in \Theta} 
    \frac{e^{\beta \sum^{\ }_{\langle s,s^{\prime}_{\ } \rangle} 
          \theta^{\ }_{s} \theta^{\ }_{s^{\prime}_{\ }} /2}}
         {\sqrt{Z}} 
      g(\bftheta) 
\, \vert 0\rangle, 
\eeq
where 
$
Z 
= 
\sum^{\ }_{\bftheta \in \Theta} 
  e^{\beta \sum^{\ }_{\langle s,s^{\prime}_{\ } \rangle} 
     \theta^{\ }_{s} \theta^{\ }_{s^{\prime}_{\ }}}
$. 
Thus, all equal-time correlation functions that can be expressed in terms of 
the $\theta^{\ }_{s}$ variables are the same as those of a $2D$ classical 
Ising model with reduced nearest-neighbor coupling $J/T = \beta$, implying 
that the critical point of the latter 
$\beta^{\ }_{c} = (1/2) \ln (\sqrt{2} + 1) \simeq 0.4406868$ 
corresponds precisely to the critical point of our quantum system. 
Notice also that the magnetization in the original $\sigma$ spin language is 
indeed the nearest-neighbor spin-spin correlation (i.e., the energy) in the 
$\theta$ spin language, 
\bea
m(\beta) 
&=& 
\frac{1}{N}\sum_i 
  \langle GS \vert \hat{\sigma}^{\textrm{z}}_{i} \vert GS \rangle 
\nonumber \\ 
&=& 
\frac{1}{Z} \sum^{\ }_{\bftheta \in \Theta} 
  e^{\beta \sum^{\ }_{\langle s,s^{\prime}_{\ } \rangle} 
     \theta^{\ }_{s} \theta^{\ }_{s^{\prime}_{\ }}}
    \left[ 
      \frac{1}{N} \sum_i \sigma^{\textrm{z}}_{i}\left(g(\bftheta)\right)
    \right] 
\nonumber \\ 
&=& 
\frac{1}{Z} \sum^{\ }_{\bftheta \in \Theta} 
  e^{\beta \sum^{\ }_{\langle s,s^{\prime}_{\ } \rangle} 
     \theta^{\ }_{s} \theta^{\ }_{s^{\prime}_{\ }}}
    \left[ 
      \frac{1}{N} \sum^{\ }_{\langle s,s^{\prime}_{\ } \rangle}
        \theta^{\ }_{s}\theta^{\ }_{s^{\prime}_{\ }} 
    \right] 
\nonumber \\ 
&=& 
\frac{1}{N} \,E_{\rm Ising}(\beta)
\;. 
\eea
Therefore, one concludes that the magnetization $m(\beta)$ is continuous 
across the transition at $\beta_c$ but there is a singularity in 
its first derivative 
\beq 
\frac{\partial m}{\partial \beta} 
= 
\frac{1}{N}\,\frac{\partial E_{\rm Ising}}{\partial \beta}
= 
-\beta^2\;\frac{1}{N}\,C_{\rm Ising}(\beta) 
\;, 
\eeq 
as the Ising model heat capacity $C_{\rm Ising}$ diverges logarithmically 
at $\beta_c$. 

In the following, we will show how such continuous phase transition is 
accompanied by a sudden, discontinuous vanishing of the topological entropy 
of the system. 

The case of a configuration of the form $hgk$, with $h \in G^{\ }_{A}$ and 
$k \in G^{\ }_{B}$ requires a few additional steps. First of all, notice that 
the composition of any two elements $g,\tilde{g} \in G$ is represented 
in the $\theta$ spin language by the site-by-site product of  the two 
configurations corresponding to $g$ and $\tilde{g}$, respectively: 
$
\theta^{\ }_{s}(g \tilde{g}) 
= 
\theta^{\ }_{s}(g) \, \theta^{\ }_{s}(\tilde{g})
$. 
In particular, 
$\theta^{\ }_{s}(hgk) = \theta^{\ }_{s}(hk) \, \theta^{\ }_{s}(g)$. 

Moreover, using similar arguments as in Ref.~\onlinecite{Castelnovo2006}, 
the star operators of a bipartite system $(A,B)$ can be 
divided into \emph{bulk} star operators, i.e., those acting 
solely on subsystem $A$ or subsystem $B$, and \emph{boundary} star operators 
acting simultaneously on $A$ and $B$ spins. The boundary star operators 
can be further subdivided into different sets according to the different 
boundaries around each connected component of $A$ and $B$ (for a total 
of $m^{\ }_{A}+m^{\ }_{B}-1$ boundaries, $m^{\ }_{A}$ and $m^{\ }_{B}$ being 
the number of connected components of $A$ and $B$, respectively). 

Let us define a \emph{collective operation} as the product of all the star 
operators in one of these sets. 
That is, the product of all the stars around a connected boundary of the
bipartition $(A,B)$. 
Clearly, the number of such collective operations is given by the number of 
sets, $m^{\ }_{A}+m^{\ }_{B}-1$. 

One can show that the subgroup $G^{\ }_{A}G^{\ }_{B} \subset G$, to which 
the product $hk$ belongs, can be generated by all the bulk star operators 
together with \emph{all but one} of the collective operators (all but one, 
independently of which one is chosen to be left out, is required because 
the product of \emph{all} boundary star operators is equivalent to the 
product of all bulk star operators). 
For example, $G^{\ }_{A}G^{\ }_{B}$ is generated by the bulk star operators 
alone in bipartitions~2 and~3 in Fig.~\ref{fig: topological partitions}, 
while the product of all boundary star operators along one of the two 
boundaries must be included to generate $G^{\ }_{A}G^{\ }_{B}$ for 
bipartitions~1 and~4. 

Let us define $\Theta^{b}_{\ } = \{ \bftheta^{b}_{\ } \}$, `$b$' for 
`bulk', to be the set of Ising spin configurations on the sites of the 
square lattice where all $\theta^{b}_{s}$ corresponding to boundary sites 
$s$ are fixed to equal $+1$. 
Let us also define $\Theta^{\delta}_{\ } = \{ \bftheta^{\delta}_{\ } \}$, 
`$\delta$' for `boundary',  
to be the set of Ising configurations where $\theta^{\delta}_{s} = +1$ for 
all bulk star operators, $\theta^{\delta}_{s} = +1$ for all boundary 
star operators belonging to \emph{one chosen boundary}, and 
$\theta^{\delta}_{s} = \pm 1$ for the remaining boundary 
star operators, so long as \emph{all $\theta^{\delta}_{s}$ spins belonging 
to the same boundary have the same sign}. 
Notice that $\Theta^{\delta}_{\ } = \{ \openone \}$ for bipartitions~2 and~3 
in Fig.~\ref{fig: topological partitions}, where $\openone$ is the 
configuration with all the spins $\theta^{\delta}_{s} = +1$. 
One can finally show that there is a one-to-one correspondence between 
the elements of $G^{\ }_{A}G^{\ }_{B}$ and the Ising configurations in 
$
\{ 
\bftheta^{b}_{\ } \bftheta^{\delta}_{\ }, 
\; 
\forall\,
\bftheta^{b}_{\ }\in\Theta^{b}_{\ }, \: 
\bftheta^{\delta}_{\ }\in\Theta^{\delta}_{\ }
\}
$, 
where $\bftheta^{b}_{\ } \bftheta^{\delta}_{\ }$ represents the site-by-site 
product of the two configurations (i.e., 
$
(\theta^{b}_{\ }\theta^{\delta}_{\ })^{\ }_{s} 
= 
\theta^{b}_{s} \theta^{\delta}_{s}
$). 
Therefore, 
\bea
\sum^{\ }_{h \in G^{\ }_{A},\, k \in G^{\ }_{B}} 
  e^{\beta \sum^{\ }_{i} \sigma^{\textrm{z}}_{i}(h g k)} 
&\equiv& 
\nonumber \\ 
&& 
\!\!\!\!\!\!\!\!\!\!\!\!\!\!\!\!\!\!\!\!\!\!\!\!\!\!
\!\!\!\!\!\!\!\!\!\!\!\!\!\!\!\!\!\!\!\!\!\!\!\!\!\!
\!\!\!\!\!
\equiv 
\sum^{\ }_{\bftheta^{b}_{\ }\in\Theta^{b}_{\ },\,
           \bftheta^{\delta}_{\ }\in\Theta^{\delta}_{\ }} 
  e^{\beta \sum^{\ }_{\langle s,s^{\prime}_{\ } \rangle} 
    \theta^{b}_{s} \theta^{\delta}_{s} \theta^{\ }_{s}(g)
    \theta^{\ }_{s^{\prime}_{\ }}(g) \theta^{\delta}_{s^{\prime}_{\ }} 
    \theta^{b}_{s^{\prime}_{\ }} }, 
\nonumber \\ 
&& 
\label{eq: log arg 2 and 3}
\eea
and in particular, 
\bea
\sum^{\ }_{\langle s,s^{\prime}_{\ } \rangle} 
  \theta^{b}_{s} \theta^{\delta}_{s} \theta^{\ }_{s}(g)
  \theta^{\ }_{s^{\prime}_{\ }}(g) \theta^{\delta}_{s^{\prime}_{\ }} 
  \theta^{b}_{s^{\prime}_{\ }} 
&=& 
\nonumber \\ 
&& 
\!\!\!\!\!\!\!\!\!\!\!\!\!\!\!\!\!\!\!\!\!\!\!\!\!\!
\!\!\!\!\!\!\!\!\!\!\!\!\!\!\!\!\!\!\!\!\!\!\!\!\!\!
= 
\sum^{s,s^{\prime}_{\ } \:\textrm{bulk}}_{\langle s,s^{\prime}_{\ } \rangle} 
  \theta^{b}_{s} \theta^{\ }_{s}(g)
  \theta^{\ }_{s^{\prime}_{\ }}(g) 
  \theta^{b}_{s^{\prime}_{\ }} 
\label{eq: theta energy 1} 
\\ 
&& 
\!\!\!\!\!\!\!\!\!\!\!\!\!\!\!\!\!\!\!\!\!\!\!\!\!\!
\!\!\!\!\!\!\!\!\!\!\!\!\!\!\!\!\!\!\!\!\!\!\!\!\!\!
+
\mathop{\sum^{s\:\textrm{bulk}}_{\langle s,s^{\prime}_{\ }\rangle}}^{s^{\prime}_{\ }\:\textrm{boundary}}
  \theta^{b}_{s} \theta^{\ }_{s}(g)
  \theta^{\ }_{s^{\prime}_{\ }}(g) \theta^{\delta}_{s^{\prime}_{\ }} 
\label{eq: theta energy 2}
\\ 
&& 
\!\!\!\!\!\!\!\!\!\!\!\!\!\!\!\!\!\!\!\!\!\!\!\!\!\!
\!\!\!\!\!\!\!\!\!\!\!\!\!\!\!\!\!\!\!\!\!\!\!\!\!\!
+ 
\sum^{s,s^{\prime}_{\ } \:\textrm{different boundaries}}_{\langle s,s^{\prime}_{\ } \rangle} 
  \theta^{\delta}_{s} \theta^{\ }_{s}(g)
  \theta^{\ }_{s^{\prime}_{\ }}(g) \theta^{\delta}_{s^{\prime}_{\ }} 
\label{eq: theta energy 3}
\\ 
&& 
\!\!\!\!\!\!\!\!\!\!\!\!\!\!\!\!\!\!\!\!\!\!\!\!\!\!
\!\!\!\!\!\!\!\!\!\!\!\!\!\!\!\!\!\!\!\!\!\!\!\!\!\!
+ 
\sum^{s,s^{\prime}_{\ } \:\textrm{same boundary}}_{\langle s,s^{\prime}_{\ } \rangle} 
  \theta^{\ }_{s}(g) 
  \theta^{\ }_{s^{\prime}_{\ }}(g), 
\label{eq: theta energy 4}
\eea
where we used the fact that if $s$ is in the bulk then 
$\theta^{\delta}_{s}=+1$, if $s$ belongs to a boundary then 
$\theta^{b}_{s}=+1$, and if both $s$ and $s^{\prime}_{\ }$ belong to the 
\emph{same} boundary then 
$\theta^{\delta}_{s} \theta^{\delta}_{s^{\prime}_{\ }} = +1$. 

Let us focus on the bipartitions of interest to compute the topological 
entropy~(\ref{eq: topo entropy}). 
First of all, in the limit 
$r,R \to \infty$ there are no nearest-neighboring stars $s$ and 
$s^{\prime}_{\ }$ belonging to two different boundaries. Therefore, the 
term~(\ref{eq: theta energy 3}) vanishes identically. 
For bipartitions~2 and~3, $\Theta^{\delta}_{\ } = \{ \openone \}$ and 
\begin{widetext}
\bea
\mathop{\sum^{\ }_{h \in G^{\ }_{A}}}^{\ }_{k \in G^{\ }_{B}} 
  e^{\beta \sum^{\ }_{i} \sigma^{\textrm{z}}_{i}(h g k)} 
&\equiv& 
e^{\beta \sum^{s,s^{\prime}_{\ }\;\textrm{boundary}}_{\langle s,s^{\prime}_{\ } \rangle} 
  \theta^{\ }_{s}(g) \theta^{\ }_{s^{\prime}_{\ }}(g)} 
\sum^{\ }_{\bftheta^{b}_{\ }\in\Theta^{b}_{\ }} 
  e^{\beta 
       \sum^{s,s^{\prime}_{\ }\:\textrm{bulk}}_{\langle s,s^{\prime}_{\ } \rangle} 
         \theta^{b}_{s} \theta^{\ }_{s}(g)
           \theta^{\ }_{s^{\prime}_{\ }}(g) \theta^{b}_{s^{\prime}_{\ }}
    }
  e^{\beta 
       \sum^{s^{\prime}_{\ }\:\textrm{boundary},s\:\textrm{bulk}}_{\langle s,s^{\prime}_{\ } \rangle} 
         \theta^{b}_{s} \theta^{\ }_{s}(g)
           \theta^{\ }_{s^{\prime}_{\ }}(g) 
    }. 
\eea
\end{widetext} 
The r.h.s. of the above equation can be interpreted as the partition 
function of an Ising model with nearest-neighbor interactions, where only 
the bulk degrees of freedom are allowed to flip starting from a given 
configuration $\bftheta(g)$. 
Clearly such partition function is invariant upon changing the initial 
configuration as long as the new one is in the same ergodic sector. 
~\\

\noindent
For example, one can equivalently choose 
\beq
\tilde{\bftheta}(g) 
= 
\left\{
  \begin{array}{ll}
    +1 & \textrm{if $s$ belongs to the bulk} \\ 
    \theta^{\ }_{s}(g) & \textrm{if $s$ belongs to the boundary}, 
  \end{array}
\right. 
\eeq
and the expression above simplifies to 
\begin{widetext}
\bea
\mathop{\sum^{\ }_{h \in G^{\ }_{A}}}^{\ }_{k \in G^{\ }_{B}} 
  e^{\beta \sum^{\ }_{i} \sigma^{\textrm{z}}_{i}(h g k)} 
&\equiv& 
e^{\beta \sum^{s,s^{\prime}_{\ }\;\textrm{boundary}}_{\langle s,s^{\prime}_{\ } \rangle} 
  \theta^{\ }_{s}(g) \theta^{\ }_{s^{\prime}_{\ }}(g)} 
\sum^{\ }_{\bftheta^{b}_{\ }\in\Theta^{b}_{\ }} 
  e^{\beta 
       \sum^{s,s^{\prime}_{\ }\:\textrm{bulk}}_{\langle s,s^{\prime}_{\ } \rangle} 
         \theta^{b}_{s} \theta^{b}_{s^{\prime}_{\ }}
    }
  e^{\beta 
       \sum^{s^{\prime}_{\ }\:\textrm{boundary},s\:\textrm{bulk}}_{\langle s,s^{\prime}_{\ } \rangle} 
         \theta^{b}_{s} \theta^{\ }_{s^{\prime}_{\ }}(g) 
    }
= 
Z^{\partial}_{2,3}(g). 
\label{eq: 2-3 contribs}
\eea
\end{widetext}
Here $Z^{\partial}_{2,3}(g)$ represents the partition function of an Ising 
model with nearest-neighbor interaction of reduced strength $J/T = \beta$, 
and with fixed spins along the boundary of bipartitions $2$ and $3$, 
respectively. 
The values of the spins at the boundary are determined by $g$. 

For bipartitions~1 and~4, 
$\Theta^{\delta}_{\ } = \{ \openone, \mathbf{f} \}$, 
where the configuration $\mathbf{f}$ has all the spins equal to $+1$ except 
for those belonging to the chosen boundary, say boundary 2 in 
Fig.~\ref{fig: topological partitions}, 
which are equal to $-1$. In this case 
\begin{widetext} 
\bea
\mathop{\sum^{\ }_{h \in G^{\ }_{A}}}^{\ }_{k \in G^{\ }_{B}} 
  e^{\beta \sum^{\ }_{i} \sigma^{\textrm{z}}_{i}(h g k)} 
&\equiv& 
e^{\beta \sum^{s,s^{\prime}_{\ }\;\textrm{same boundary}}_{\langle s,s^{\prime}_{\ } \rangle} 
  \theta^{\ }_{s}(g) \theta^{\ }_{s^{\prime}_{\ }}(g)} 
\sum^{\ }_{\bftheta^{b}_{\ }\in\Theta^{b}_{\ }} 
  e^{\beta 
       \sum^{s,s^{\prime}_{\ }\:\textrm{bulk}}_{\langle s,s^{\prime}_{\ } \rangle} 
         \theta^{b}_{s} \theta^{b}_{s^{\prime}_{\ }}
    }
e^{\beta 
     \sum^{s^{\prime}_{\ }\:\textrm{boundary 1},s\:\textrm{bulk}}_{\langle s,s^{\prime}_{\ } \rangle} 
       \theta^{b}_{s} \theta^{\ }_{s^{\prime}_{\ }}(g) 
  } 
\nonumber \\ 
&\times& 
\left( 
  e^{\beta 
       \sum^{s^{\prime}_{\ }\:\textrm{boundary 2},s\:\textrm{bulk}}_{\langle s,s^{\prime}_{\ } \rangle} 
         \theta^{b}_{s} \theta^{\ }_{s^{\prime}_{\ }}(g) 
    } 
  + 
  e^{-\beta 
       \sum^{s^{\prime}_{\ }\:\textrm{boundary 2},s\:\textrm{bulk}}_{\langle s,s^{\prime}_{\ } \rangle} 
         \theta^{b}_{s} \theta^{\ }_{s^{\prime}_{\ }}(g) 
    } 
\right) 
\nonumber \\ 
&=&
Z^{\partial}_{1,4}(g) + Z^{\partial,\,\textrm{twisted}}_{1,4}(g). 
\label{eq: 1-4 contribs}
\eea
\end{widetext} 
Here $Z^{\partial}_{1,4}(g)$ are the analog of $Z^{\partial}_{2,3}(g)$ for 
bipartitions $1$ and $4$, respectively, 
while $Z^{\partial,\,\textrm{twisted}}_{1,4}(g)$ differ from the 
former by the fact that all the (fixed) spins belonging to boundary 2 in 
bipartitions $1$ and $4$ respectively have been flipped. 
In other words, $Z^{\partial}_{1,4}(g)$ represents the partition function 
of an Ising model with nearest-neighbor interaction of reduced strength 
$J/T = \beta$, and with fixed spins along the boundary of bipartitions 
$1$ and $4$, respectively. 
The partition functions $Z^{\partial,\,\textrm{twisted}}_{1,4}(g)$ differ 
in that the spins along one of the two boundaries have been flipped with 
respect to their values in $Z^{\partial}_{1,4}(g)$. 
Again, the values of the spins at the boundary are determined by $g$. 

In this notation, the topological entropy of the system can be written as 
\begin{widetext}
\bea
S^{\ }_{\textrm{topo}} 
&=&
\lim^{\ }_{r, R \to \infty} 
\left\{\vphantom{\sum} 
\frac{1}{Z} 
  \sum^{\ }_{g \in G} e^{\beta \sum^{\ }_{i} \sigma^{\textrm{z}}_{i}(g)} 
  \log^{\ }_{2} 
    \frac{\left[ Z^{\partial}_{1}(g) + 
                 Z^{\partial,\,\textrm{twisted}}_{1}(g) \right]
          \left[ Z^{\partial}_{4}(g) + 
	         Z^{\partial,\,\textrm{twisted}}_{4}(g) \right]}
         {Z^{\partial}_{2}(g) Z^{\partial}_{3}(g)}
\vphantom{\sum^{\ }_{g\in G}}\right\}, 
\nonumber \\
\label{eq: formula for the topo entropy}
\eea
\end{widetext}
where the sum over $g$ acts as a weighed average of the logarithmic 
term over all possible values of the spins at the boundary. 
Notice that in Eq.~(\ref{eq: formula for the topo entropy}) the partitions 
with two boundaries, and hence with non-trivial topology, are those that 
appear with two contributions (bipartitions 1 and 4), corresponding to some 
relative boundary conditions (BCs) and their twisted counterparts. 
These contributions, as we show below in detail, are responsible 
for the non-vanishing topological entropy. 
In the topological phase, the two partition 
functions for the twisted and untwisted BCs contribute equally, and in 
the non-topological phase, one partition function is exponentially 
suppressed when compared to the other, in the thermodynamic limit. 
Therefore, there is an extra entropy contribution in one of the phases 
depending on whether the boundaries of topologically non-trivial 
bipartitions are twisted or not relative to one another. 

{}From Eq.~(\ref{eq: formula for the topo entropy}), the behavior of the 
topological entropy can be qualitatively argued as follows. 
Deep in the disordered phase, where the correlations are short ranged, 
the choice of boundary conditions is likely to affect the partition function 
of the system only with exponentially small corrections. Thus, we can 
expect to have 
$
Z^{\partial}_{1}(g)Z^{\partial}_{4}(g) 
\simeq 
Z^{\partial,\,\textrm{twisted}}_{1}(g)Z^{\partial}_{4}(g) 
\simeq 
\,\ldots\, 
\simeq 
Z^{\partial}_{2}(g) Z^{\partial}_{3}(g)
$
and $S^{\ }_{\textrm{topo}} = 2$. 
On the other hand, deep in the (ferromagnetically) ordered phase the 
partition function of a system with twisted boundary conditions is 
exponentially suppressed with respect to the one without the twist. 
Thus, 
$
Z^{\partial}_{1}(g) 
\gg 
Z^{\partial,\,\textrm{twisted}}_{1}(g) 
$, 
$
Z^{\partial}_{4}(g) 
\gg 
Z^{\partial,\,\textrm{twisted}}_{4}(g) 
$, 
while 
$
Z^{\partial}_{1}(g)Z^{\partial}_{4}(g) 
\simeq 
Z^{\partial}_{2}(g) Z^{\partial}_{3}(g)
$ 
still holds. 
This leads to $S^{\ }_{\textrm{topo}} = 0$. 

In the following two sections we will show with rigorous arguments that 
the behavior of 
the topological entropy across the transition is strongly first order, 
with a sudden jump from $S^{\ }_{\textrm{topo}} = 2$ to 
$S^{\ }_{\textrm{topo}} = 0$. 
%
%

\subsubsection{
The disordered phase ($\beta < \beta^{\ }_{c}$)
              }
In the limit of small $\beta$, namely above the ordering transition, one can 
compute $S^{\ }_{\textrm{topo}}$ via the high-temperature expansion of the 
Ising model with fixed spins at the boundary. 

Let us rewrite, 
\bea
Z^{\partial}_{2}(g) 
&=& 
\sum^{\ }_{\bftheta^{b}_{\ }\in\Theta^{b}_{\ }} \: 
  \prod^{\ }_{\textrm{bonds}\,i} 
    \left(
      \cosh\beta 
      + 
      \sinh\beta \: \theta^{\ }_{s^{\ }_{i}} \theta^{\ }_{s^{\prime}_{i}}
    \right)
\nonumber \\ 
&=& 
2^{N/2-\delta}_{\ } c^{N}_{\ } \: 
  \sum^{\ }_{\mathcal{G}} \, 
    t^{\ell(\mathcal{G})}_{\ } \, 
      \prod^{\ }_{s \in \mathcal{E}_{\!\mathcal{G}}} \theta^{\ }_{s}(g), 
\nonumber \\ 
\eea
where $s^{\ }_{i},s^{\prime}_{i}$ are the sites at the ends of bond $i$, 
$N$ is the total number of bonds on the lattice, 
$\delta$ is the length of the boundary in number of $\theta$ spins, 
$c = \cosh\beta$, $t = \tanh\beta$. 
The sum over $\mathcal{G}$ runs over all possible graphs on the bonds of the 
square lattice, composed entirely of closed loops and open strings connecting 
two boundary spins. 
The product 
$\prod^{\ }_{s \in \mathcal{E}_{\!\mathcal{G}}} \theta^{\ }_{s}(g)$
encompasses all the boundary spins that appear as end points 
(the set $\mathcal{E}_{\!\mathcal{G}}$) of open strings in $\mathcal{G}$. 
Finally, $\ell(\mathcal{G})$ is the total length of the closed loops and open 
strings in $\mathcal{G}$. 

Analogously for $Z^{\partial}_{3}(g)$. 
The case of $Z^{\partial}_{1}(g)$ and $Z^{\partial}_{4}(g)$ 
differs from $Z^{\partial}_{2}(g)$ and $Z^{\partial}_{3}(g)$ in that there 
are now two types of open strings: those going from one 
boundary to itself and those connecting the two boundaries (see 
Fig.~\ref{fig: topological partitions}). 

Next, let us compare the product $Z^{\partial}_{2}(g) Z^{\partial}_{3}(g)$ 
with the product $Z^{\partial}_{1}(g) Z^{\partial}_{4}(g)$. 
Notice that bipartitions 2 and 3 have precisely the same total combined 
boundary as bipartitions 1 and 4. In order for a graph to appear in one of 
the two products and not in the other, it needs to comprise loops or strings 
that are able to tell the difference between to two possible origins 
(2 + 3 vs. 1 + 4) of the total combined boundary. 

\begin{figure}[ht]
\vspace{-0.2 cm}
\includegraphics[width=0.9\columnwidth]{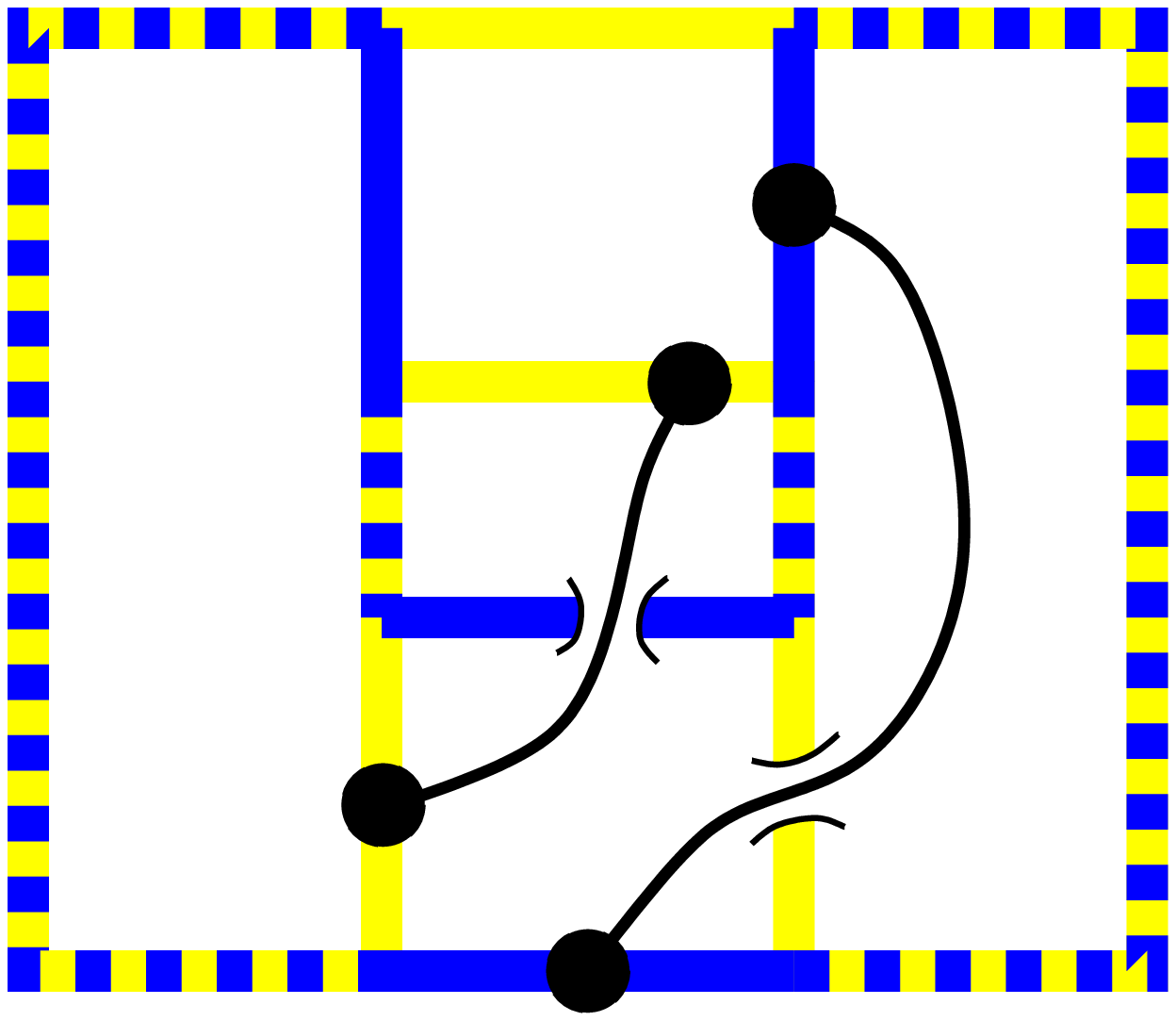}
\\ 
\includegraphics[width=0.9\columnwidth]{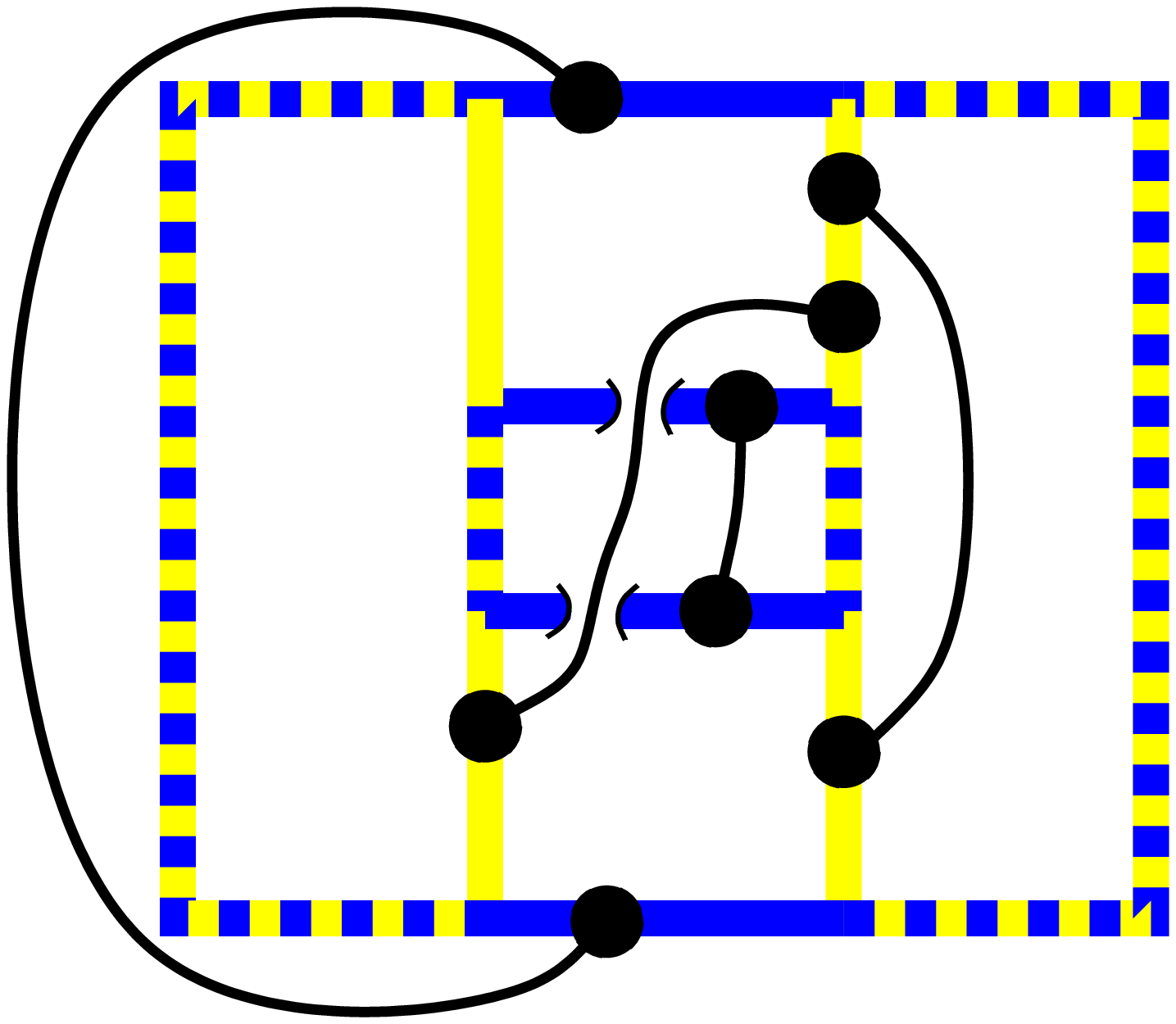}
\caption{
\label{fig: loop product}
(Color online) -- 
Examples of open strings that appear in the expansion of 
$Z^{\partial}_{2}(g) Z^{\partial}_{3}(g)$ but are not present in the 
expansion of $Z^{\partial}_{1}(g) Z^{\partial}_{4}(g)$ (Top), 
and vice versa (Bottom). 
Top panel: 
The thick yellow and blue lines correspond to the boundaries in bipartitions 
2 and 3 respectively. 
Boundaries belonging to both are shown in a thick dashed yellow-blue 
pattern. 
The strings in question are symbolically represented by thin black lines. 
Bottom panel: 
Same color coding, with yellow corresponding to bipartition 4 and blue 
corresponding to biparition 1. 
Notice that strings appearing in one expansion and not in the other must 
connect boundaries of the same solid color, and therefore cannot be 
shorter than $R-2r$. 
}
\end{figure}

Examples of such open strings are shown in Fig.~\ref{fig: loop product}. 
One can show that these telltale strings, and the analogous closed loops, 
cannot be arbitrarily short, and their length 
is bounded from below by $R-2r$. 
As a consequence, the corresponding graphs are exponentially suppressed 
at least as $t^{R-2r}_{\ }$, and in the limit $r,R \to \infty$ 
with $R-2r \to \infty$, implicit in the definition of the 
topological entropy, one obtains 
\beq
\frac{Z^{\partial}_{1}(g) Z^{\partial}_{4}(g)}
{Z^{\partial}_{2}(g) Z^{\partial}_{3}(g)}
\to 1 
. 
\label{eq: high T lim 1}
\eeq

Similar considerations apply when comparing the product 
$Z^{\partial}_{1}(g) Z^{\partial}_{4}(g)$ with products of the kind 
$Z^{\partial,\,\textrm{twisted}}_{1}(g) Z^{\partial}_{4}(g)$. 
In this case, the boundaries involved are exaclty the same, and the relevant 
telltale elements of the graph are open strings connecting one of the two 
components of the boundary with the other. Such strings are in fact the only 
elements that are sensitive to the twisted boundary conditions. 
Clearly the length of these strings is bounded from below by $R-2r$, 
and 
\beq
\frac{Z^{\partial,\,\textrm{twisted}}_{1}(g) Z^{\partial}_{4}(g)}
{Z^{\partial}_{1}(g) Z^{\partial}_{4}(g)}
\to 1
\label{eq: high T lim 2}
\eeq
exponentially fast, at least as $t^{R-2r}$, with $(R-2r)\to\infty$. 

Of course, our reasoning is correct up to the point where the high 
temperature expansion breaks down, and entropic contributions balance 
the exponential suppression. Said differently, this is the case when the
correlation length in the Ising model goes to infinity, and the large
$r,R$ limit does not guarantee that the ratios of products of
partition functions above tend to one. 

Given Eqs.~(\ref{eq: high T lim 1},\ref{eq: high T lim 2}), we can finally 
use Eq.~(\ref{eq: formula for the topo entropy}) to obtain the topological 
entropy of the system throughout the disordered phase 
$\beta < \beta^{\ }_{c} \simeq 0.4406868$, 
\bea
S^{\ }_{\textrm{topo}} 
&=&
\frac{1}{Z} 
  \sum^{\ }_{g \in G} e^{\beta \sum^{\ }_{i} \sigma^{\textrm{z}}_{i}(g)} 
  \log^{\ }_{2} 4 
\nonumber \\ 
&=& 
\log^{\ }_{2} 4 
= 2 
. 
\nonumber
\eea
%
%
%

\subsubsection{
The Landau-Ginzburg ordered phase ($\beta > \beta^{\ }_{c}$)
              }
What happens below this transition? 
Rather than attempting a low-temperature expansion, it is convenient to use 
the duality relations derived by A.~Bugrij and V.~Shadura 
in Ref.~\onlinecite{Bugrij1996} for the inhomogeneous, finite-size Ising 
model. 
In particular, they obtained the duality relations for a square lattice 
Ising model wrapped around a cylinder of finite length, with fixed, free 
and mixed boundary conditions. 
Following the usual convention, let us label $\tilde{\beta}$ the coupling 
constant of the dual Ising model (defined on the plaquettes of the original 
lattice), which is related to $\beta$ by the duality relation 
$\sinh \beta \, \sinh \tilde{\beta} = 1$. 
Let us also indicate with $Z(\delta,\delta^{\prime}_{\ })$ and 
$\tilde{Z}(\delta,\delta^{\prime}_{\ })$ the partition functions of the 
system on the finite cylinder and its dual, with 
$\delta$, $\delta^{\prime}_{\ }$ specifying the boundary conditions, namely 
$\delta,\,\delta^{\prime}_{\ } = \Circle,\,\times$ for free and fixed 
boundary spins, respectively. 
With this notation in mind, the results by Bugrij and Shadura -- to the 
purpose of the present paper -- can be summarized by~\cite{Bugrij1996} 
\bea
\tilde{Z}(\Circle,\Circle) 
&=& 
\mathcal{K} \left[ Z(\times,\times) 
+ 
Z^{\textrm{twisted}}_{\ }(\times,\times) \right] 
\label{eq: ++ +- boundary}
\\ 
\tilde{Z}(\times,\Circle) &=& \mathcal{K}\, Z(\Circle,\times) 
\label{eq: 0+ boundary}
\\ 
\tilde{Z}(\Circle,\times) &=& \mathcal{K}\, Z(\times,\Circle), 
\label{eq: +0 boundary}
\eea
where $Z^{\textrm{twisted}}_{\ }(\times,\times)$ differs from 
$Z(\times,\times)$ by the fact that the fixed boundary spins at one 
end of the cylinder have been flipped. 
Notice that the proportionality coefficient $\mathcal{K}$ is the same in all 
the equations, and that $Z(\Circle,\times) = Z(\times,\Circle)$ and 
$\tilde{Z}(\times,\Circle) = \tilde{Z}(\Circle,\times)$. 

Let us then consider $Z^{\partial}_{2}$ in Eq.~(\ref{eq: 2-3 contribs}). 
Thanks to the nearest-neighbor character of the interaction between $\theta$ 
spins, subsystem $A$ interacts only with itself and with the boundary 
$\delta^{\ }_{2}$, and so does subsystem $B$. 
Thus, one can factorize the two subsystems and obtain 
(cfr. Eq.(\ref{eq: 2-3 contribs})) 
\bea
\mathop{\sum^{\ }_{h \in G^{\ }_{2A}}}^{\ }_{k \in G^{\ }_{2B}} 
  e^{\beta \sum^{\ }_{i} \sigma^{\textrm{z}}_{i}(h g k)} 
&\equiv& 
e^{\beta \sum^{s,s^{\prime}_{\ }\in\delta^{\ }_{2}}
              _{\langle s,s^{\prime}_{\ } \rangle} 
  \theta^{\ }_{s}(g) \theta^{\ }_{s^{\prime}_{\ }}(g)} 
\nonumber \\ 
&& 
\!\!\!\!\!\!\!\!\!\!\!\!\!\!\!\!\!\!\!\!\!\!\!\!\!\!\!\!\!\!
\!\!\!\!\!\!\!\!\!\!\!\!\!\!\!\!\!\!\!\!\!\!\!\!\!\!\!\!\!\!
\times 
\left( 
\sum^{\ }_{\bftheta^{b}_{2A}\in\Theta^{b}_{2A}} 
  e^{\beta 
       \sum^{s,s^{\prime}_{\ }\:\textrm{bulk}}
           _{\langle s,s^{\prime}_{\ } \rangle} 
         \theta^{b}_{s} \theta^{b}_{s^{\prime}_{\ }}
    }
  e^{\beta 
       \sum^{s^{\prime}_{\ }\:\textrm{boundary},s\:\textrm{bulk}}_{\langle s,s^{\prime}_{\ } \rangle} 
         \theta^{b}_{s} \theta^{\ }_{s^{\prime}_{\ }}(g) 
    }
\right) 
\nonumber \\ 
&&
\!\!\!\!\!\!\!\!\!\!\!\!\!\!\!\!\!\!\!\!\!\!\!\!\!\!\!\!\!\!
\!\!\!\!\!\!\!\!\!\!\!\!\!\!\!\!\!\!\!\!\!\!\!\!\!\!\!\!\!\!
\times 
\left( 
\sum^{\ }_{\bftheta^{b}_{2B}\in\Theta^{b}_{2B}} 
  e^{\beta 
       \sum^{s,s^{\prime}_{\ }\:\textrm{bulk}}
           _{\langle s,s^{\prime}_{\ } \rangle} 
         \theta^{b}_{s} \theta^{b}_{s^{\prime}_{\ }}
    }
  e^{\beta 
       \sum^{s^{\prime}_{\ }\:\textrm{boundary},s\:\textrm{bulk}}
           _{\langle s,s^{\prime}_{\ } \rangle} 
         \theta^{b}_{s} \theta^{\ }_{s^{\prime}_{\ }}(g) 
    }
\right) 
\nonumber \\ 
&& 
\!\!\!\!\!\!\!\!\!\!\!\!\!\!\!\!\!\!\!\!\!\!\!\!\!\!\!\!\!\!
= 
\left( 
  e^{\beta \sum^{s,s^{\prime}_{\ }\in\delta^{\ }_{2}}
                _{\langle s,s^{\prime}_{\ } \rangle} 
    \theta^{\ }_{s}(g) \theta^{\ }_{s^{\prime}_{\ }}(g)} 
\right) 
Z^{\partial}_{2A}(g) \, Z^{\partial}_{2B}(g). 
\nonumber \\ 
\label{eq: 2 factorized}
\eea
Similar arguments apply to bipartition 3, 
\bea
\mathop{\sum^{\ }_{h \in G^{\ }_{3A}}}^{\ }_{k \in G^{\ }_{3B}} 
  e^{\beta \sum^{\ }_{i} \sigma^{\textrm{z}}_{i}(h g k)} 
&=& 
\nonumber \\
&& 
\!\!\!\!\!\!\!\!\!\!\!\!\!\!\!\!\!\!\!\!\!\!\!\!\!\!\!\!\!\!
= 
\left( 
  e^{\beta \sum^{s,s^{\prime}_{\ } \in \delta^{\ }_{3}}
                _{\langle s,s^{\prime}_{\ } \rangle} 
    \theta^{\ }_{s}(g) \theta^{\ }_{s^{\prime}_{\ }}(g)} 
\right) 
Z^{\partial}_{3A}(g) \, Z^{\partial}_{3B}(g), 
\nonumber \\ 
\label{eq: 3 factorized}
\eea
and with a few more steps, to bipartitions 1 and 4 as well 
(cfr. Eq.(\ref{eq: 1-4 contribs})), 
\bea
\mathop{\sum^{\ }_{h \in G^{\ }_{1A}}}^{\ }_{k \in G^{\ }_{1B}} 
  e^{\beta \sum^{\ }_{i} \sigma^{\textrm{z}}_{i}(h g k)} 
&=& 
\nonumber \\
&& 
\!\!\!\!\!\!\!\!\!\!\!\!\!\!\!\!\!\!\!\!\!\!\!\!\!\!\!\!\!\!
= 
\left( 
e^{\beta \sum^{s,s^{\prime}_{\ } \in \delta^{\ }_{1},\,\textrm{boundary 1}}
               _{\langle s,s^{\prime}_{\ } \rangle} 
    \theta^{\ }_{s}(g) \theta^{\ }_{s^{\prime}_{\ }}(g)} 
\right) 
\nonumber \\ 
&& 
\!\!\!\!\!\!\!\!\!\!\!\!\!\!\!\!\!\!\!\!\!\!\!\!\!\!\!\!\!\!
\times
\left( 
e^{\beta \sum^{s,s^{\prime}_{\ } \in \delta^{\ }_{1},\,\textrm{boundary 2}}
                _{\langle s,s^{\prime}_{\ } \rangle} 
    \theta^{\ }_{s}(g) \theta^{\ }_{s^{\prime}_{\ }}(g)} 
\right) 
\nonumber \\ 
&& 
\!\!\!\!\!\!\!\!\!\!\!\!\!\!\!\!\!\!\!\!\!\!\!\!\!\!\!\!\!\!
\times
\left[ 
  Z^{\partial}_{1A}(g) 
  + 
  Z^{\partial,\,\textrm{twisted}}_{1A}(g)
\right] 
\, 
Z^{\partial}_{1B^{\ }_{1}}(g) 
\, 
Z^{\partial}_{1B^{\ }_{2}}(g), 
\nonumber \\ 
\label{eq: 1 factorized}
\eea
and 
\bea
\mathop{\sum^{\ }_{h \in G^{\ }_{4A}}}^{\ }_{k \in G^{\ }_{4B}} 
  e^{\beta \sum^{\ }_{i} \sigma^{\textrm{z}}_{i}(h g k)} 
&=& 
\nonumber \\
&& 
\!\!\!\!\!\!\!\!\!\!\!\!\!\!\!\!\!\!\!\!\!\!\!\!\!\!\!\!\!\!
= 
\left( 
e^{\beta \sum^{s,s^{\prime}_{\ } \in \delta^{\ }_{4},\,\textrm{boundary 1}}
               _{\langle s,s^{\prime}_{\ } \rangle} 
    \theta^{\ }_{s}(g) \theta^{\ }_{s^{\prime}_{\ }}(g)} 
\right) 
\nonumber \\ 
&& 
\!\!\!\!\!\!\!\!\!\!\!\!\!\!\!\!\!\!\!\!\!\!\!\!\!\!\!\!\!\!
\times
\left( 
e^{\beta \sum^{s,s^{\prime}_{\ } \in \delta^{\ }_{4},\,\textrm{boundary 2}}
                _{\langle s,s^{\prime}_{\ } \rangle} 
    \theta^{\ }_{s}(g) \theta^{\ }_{s^{\prime}_{\ }}(g)} 
\right) 
\nonumber \\ 
&& 
\!\!\!\!\!\!\!\!\!\!\!\!\!\!\!\!\!\!\!\!\!\!\!\!\!\!\!\!\!\!
\times
Z^{\partial}_{4A^{\ }_{1}}(g) 
\, 
Z^{\partial}_{4A^{\ }_{2}}(g) 
\, 
\left[ 
  Z^{\partial}_{4B}(g)
  + 
  Z^{\partial,\,\textrm{twisted}}_{4B}(g)
\right], 
\nonumber \\ 
\label{eq: 4 factorized}
\eea
where $1B^{\ }_{1}$ and $1B^{\ }_{2}$ refer to the two connected components 
of subsystem $B$ in bipartition 1, i.e., the component inside boundary 1 and 
the component outside boundary 2, and analogously for $4A^{\ }_{1}$ and 
$4A^{\ }_{2}$. 

\begin{figure}[ht]
\vspace{0.2 cm}
\includegraphics[width=0.98\columnwidth]{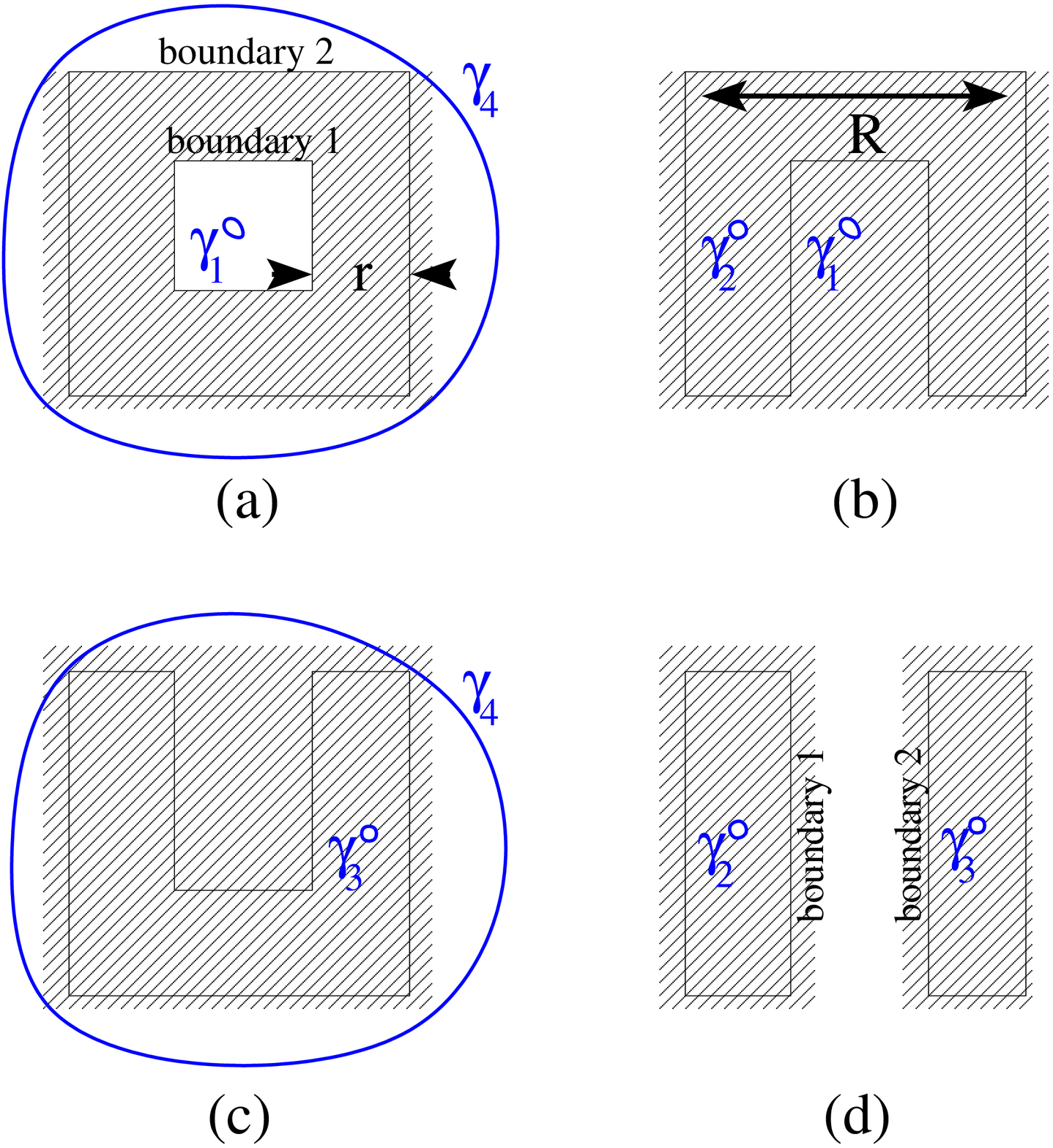}
\caption{
\label{fig: topological partitions bis}
(Color online) -- 
Illustration of the four bipartitions used to compute the topological 
entropy in Ref.~\onlinecite{Levin2006}, with a possible choice for the 
additional boundaries needed to map each partition onto an infinite 
cylinder in the thermodynamic limit. 
}
\end{figure}

In order to apply Eqs.~(\ref{eq: ++ +- boundary}-\ref{eq: +0 boundary}) 
to the present case, some further considerations on the bipartitions in 
Fig.~\ref{fig: topological partitions} are needed. 
Recall that, although $S^{\ }_{\textrm{topo}}$ is indeed a quantity of 
order one, we expressed it in Eq.~(\ref{eq: formula for the topo entropy}) 
in terms of a ratio of extensive partition functions $Z^{\partial}_{i}(g)$. 
Thus, any sub-extensive correction to these partition functions 
(i.e., $\mathcal{O}(2^{N^{\alpha}_{\ }}_{\ })$, with $\alpha < 1$, $N$ 
being the number of degrees of freedom in the system) will only 
amount to an exponentially small correction to $S^{\ }_{\textrm{topo}}$, 
that vanishes in the thermodynamic limit. 
In this context, the partition function $Z^{\partial}_{1A}(g)$ 
(see Fig.~\ref{fig: topological partitions bis}) 
is `equivalent', in the thermodynamic limit, to the partition function of 
an Ising model on an infinite cylinder with fixed boundaries at the edges 
(boundary 1 and 2, respectively). 
Similarly, the partition function $Z^{\partial}_{1B^{\ }_{1}}(g)$ can be 
regarded as that of an Ising model on an infinite cylinder with fixed 
boundary conditions on one edge (boundary 1) and open boundary conditions 
on a suitably introduced boundary $\gamma^{\ }_{1}$. 
Finally, the same approach can be used for $Z^{\partial}_{1B^{\ }_{2}}(g)$, 
with fixed boundary conditions on one edge (boundary 2) and open boundary 
conditions on another suitably introduced boundary $\gamma^{\ }_{4}$. 
Qualitatively, this is illustrated in 
Fig.~\ref{fig: topological partitions bis}a, 
where the spins on boundary 1 and boundary 2 are fixed and those belonging 
to $\gamma^{\ }_{1}$ and $\gamma^{\ }_{4}$ are free. 
Notice that the mapping onto infinite cylinders requires the distance 
between any of the $\gamma^{\ }_{i}$ boundaries introduced in 
Fig.~\ref{fig: topological partitions bis} and any of the original 
boundaries in Fig.~\ref{fig: topological partitions} to diverge with 
system size. 
Similar arguments apply to bipartitions 2, 3 and 4. 
This leads to a correspondence between our factorized partition functions 
and those used in Ref.\onlinecite{Bugrij1996}, namely 
\bea
Z^{\partial}_{1B^{\ }_{1}}(g), 
Z^{\partial}_{1B^{\ }_{2}}(g), 
Z^{\partial}_{2A}(g), 
&& 
\nonumber \\
Z^{\partial}_{2B}(g), 
Z^{\partial}_{3A}(g), 
Z^{\partial}_{3B}(g), 
&& 
\nonumber \\
Z^{\partial}_{4A^{\ }_{1}}(g), 
Z^{\partial}_{4A^{\ }_{2}}(g) 
&\sim& 
Z(\Circle,\times) 
\nonumber \\
Z^{\partial}_{1A}(g), 
Z^{\partial}_{4B}(g) 
&\sim& 
Z(\times,\times)
\nonumber \\
Z^{\partial,\,\textrm{twisted}}_{1A}(g), 
Z^{\partial,\,\textrm{twisted}}_{4B}(g) 
&\sim& 
Z^{\textrm{twisted}}_{\ }(\times,\times). 
\nonumber
\eea

The results in Ref.~\onlinecite{Bugrij1996} can then be applied to our 
systems and lead to the following equations: 
\begin{subequations} 
\bea
\tilde{Z}^{\partial}_{1A}(\Circle,\Circle) 
&\propto& 
Z^{\partial}_{1A}(g) + Z^{\partial,\,\textrm{twisted}}_{1A}(g) 
\\ 
\tilde{Z}^{\partial}_{1B^{\ }_{1}}(\times,\Circle) 
&\propto& 
Z^{\partial}_{1B^{\ }_{1}}(g) 
\\ 
\tilde{Z}^{\partial}_{1B^{\ }_{2}}(\times,\Circle) 
&\propto& 
Z^{\partial}_{1B^{\ }_{2}}(g) 
\eea
\bea
\tilde{Z}^{\partial}_{2A}(\times,\Circle) 
&\propto& 
Z^{\partial}_{2A}(g) 
\\ 
\tilde{Z}^{\partial}_{2B}(\times,\Circle) 
&\propto& 
Z^{\partial}_{2B}(g) 
\eea
\bea
\tilde{Z}^{\partial}_{3A}(\times,\Circle) 
&\propto& 
Z^{\partial}_{3A}(g) 
\\ 
\tilde{Z}^{\partial}_{3B}(\times,\Circle) 
&\propto& 
Z^{\partial}_{3B}(g) 
\eea
\bea
\tilde{Z}^{\partial}_{4B}(\Circle,\Circle) 
&\propto& 
Z^{\partial}_{4B}(g) + Z^{\partial,\,\textrm{twisted}}_{4B}(g) 
\\ 
\tilde{Z}^{\partial}_{4A^{\ }_{1}}(\times,\Circle) 
&\propto& 
Z^{\partial}_{4A^{\ }_{1}}(g) 
\\ 
\tilde{Z}^{\partial}_{4A^{\ }_{2}}(\times,\Circle) 
&\propto& 
Z^{\partial}_{4A^{\ }_{2}}(g). 
\eea
\label{eq: duality relations}
\end{subequations}

For convenience of notation, let us define the dual partition functions for 
the whole system in the different bipartitions 
\bea
\tilde{Z}^{\partial}_{1} 
&\equiv& 
\tilde{Z}^{\partial}_{1A}(\Circle,\Circle) 
\tilde{Z}^{\partial}_{1B^{\ }_{1}}(\times,\Circle) 
\tilde{Z}^{\partial}_{1B^{\ }_{2}}(\times,\Circle) 
\\ 
\tilde{Z}^{\partial}_{2} 
&\equiv& 
\tilde{Z}^{\partial}_{2A}(\times,\Circle) 
\tilde{Z}^{\partial}_{2B}(\times,\Circle) 
\\ 
\tilde{Z}^{\partial}_{3} 
&\equiv& 
\tilde{Z}^{\partial}_{3A}(\times,\Circle) 
\tilde{Z}^{\partial}_{3B}(\times,\Circle) 
\\ 
\tilde{Z}^{\partial}_{4} 
&\equiv& 
\tilde{Z}^{\partial}_{4A^{\ }_{1}}(\times,\Circle) 
\tilde{Z}^{\partial}_{4A^{\ }_{2}}(\times,\Circle) 
\tilde{Z}^{\partial}_{4B}(\Circle,\Circle) 
\eea

Finally, we have all the ingredients to evaluate the topological 
entropy for $\beta > \beta^{\ }_{c}$. 
Let us first rewrite Eq.~(\ref{eq: formula for the topo entropy}) using 
Eqs.~(\ref{eq: 2 factorized}-\ref{eq: 4 factorized}) instead of 
Eqs.~(\ref{eq: 2-3 contribs}-\ref{eq: 1-4 contribs}): 
\begin{widetext}
\bea
S^{\ }_{\textrm{topo}} 
&=&
\lim^{\ }_{r, R \to \infty} 
\left\{\vphantom{\sum^{\ }_{g\in G}} 
\frac{1}{Z} 
  \sum^{\ }_{g \in G} e^{\beta \sum^{\ }_{i} \sigma^{\textrm{z}}_{i}(g)} 
\right. 
\nonumber\\
&&
\qquad\qquad
\times 
\left[\vphantom{\sum}
  \log^{\ }_{2} 
    \frac{\left[ Z^{\partial}_{1A}(g) + 
                 Z^{\partial,\,\textrm{twisted}}_{1A}(g) \right]
          Z^{\partial}_{1B^{\ }_{1}}(g)
          Z^{\partial}_{1B^{\ }_{2}}(g)
          Z^{\partial}_{4A^{\ }_{1}}(g)
          Z^{\partial}_{4A^{\ }_{2}}(g)
	  \left[ Z^{\partial}_{4B}(g) + 
	         Z^{\partial,\,\textrm{twisted}}_{4B}(g) \right]}
         {Z^{\partial}_{2A}(g) Z^{\partial}_{2B}(g)
	  Z^{\partial}_{3A}(g) Z^{\partial}_{3B}(g)}
\right. 
\nonumber \\ 
&&
\qquad\qquad\qquad\qquad\qquad
\qquad\qquad\qquad\qquad\qquad
\left. 
\left. 
+ 
\sum^{s,s^{\prime}_{\ } \in \delta^{\ }_{1}+\delta^{\ }_{4}}
               _{\langle s,s^{\prime}_{\ } \rangle} 
    \theta^{\ }_{s}(g) \theta^{\ }_{s^{\prime}_{\ }}(g)
\; - 
\sum^{s,s^{\prime}_{\ } \in \delta^{\ }_{2}+\delta^{\ }_{3}}
               _{\langle s,s^{\prime}_{\ } \rangle} 
    \theta^{\ }_{s}(g) \theta^{\ }_{s^{\prime}_{\ }}(g)
\vphantom{\sum}\right]
\vphantom{\sum^{\ }_{g\in G}}\right\}, 
\label{eq: Stopo above beta_c}
\eea
where the last two terms inside the square brackets come from the 
exponential factors in Eqs.~(\ref{eq: 2 factorized}-\ref{eq: 4 factorized}). 
Using the duality relations~(\ref{eq: duality relations}), we can identify 
\bea
\frac{\left[ Z^{\partial}_{1A}(g) + 
             Z^{\partial,\,\textrm{twisted}}_{1A}(g) \right]
      Z^{\partial}_{1B^{\ }_{1}}(g)
      Z^{\partial}_{1B^{\ }_{2}}(g)
      Z^{\partial}_{4A^{\ }_{1}}(g)
      Z^{\partial}_{4A^{\ }_{2}}(g)
      \left[ Z^{\partial}_{4B}(g) + 
	     Z^{\partial,\,\textrm{twisted}}_{4B}(g) \right]}
     {Z^{\partial}_{2A}(g) Z^{\partial}_{2B}(g)
      Z^{\partial}_{3A}(g) Z^{\partial}_{3B}(g)}
&=& 
\nonumber \\ 
&& 
\!\!\!\!\!\!\!\!\!\!\!\!\!\!\!\!\!\!\!\!\!\!\!\!\!\!
\!\!\!\!\!\!\!\!\!\!\!\!\!\!\!\!\!\!\!\!\!\!\!\!\!\!
\!\!\!\!\!\!\!\!\!\!\!\!\!\!\!\!\!\!\!\!\!\!\!\!\!\!
\!\!\!\!\!\!\!\!\!\!\!\!\!\!\!\!\!\!\!\!\!\!\!\!\!\!
= 
\frac{\tilde{Z}^{\partial}_{1A}(\Circle,\Circle)
      \tilde{Z}^{\partial}_{1B^{\ }_{1}}(\times,\Circle)
      \tilde{Z}^{\partial}_{1B^{\ }_{2}}(\times,\Circle)
      \tilde{Z}^{\partial}_{4A^{\ }_{1}}(\times,\Circle)
      \tilde{Z}^{\partial}_{4A^{\ }_{2}}(\times,\Circle)
      \tilde{Z}^{\partial}_{4B}(\Circle,\Circle)} 
     {\tilde{Z}^{\partial}_{2A}(\times,\Circle) \tilde{Z}^{\partial}_{2B}(\times,\Circle)
      \tilde{Z}^{\partial}_{3A}(\times,\Circle) \tilde{Z}^{\partial}_{3B}(\times,\Circle)}
\nonumber \\ 
&& 
\!\!\!\!\!\!\!\!\!\!\!\!\!\!\!\!\!\!\!\!\!\!\!\!\!\!
\!\!\!\!\!\!\!\!\!\!\!\!\!\!\!\!\!\!\!\!\!\!\!\!\!\!
\!\!\!\!\!\!\!\!\!\!\!\!\!\!\!\!\!\!\!\!\!\!\!\!\!\!
\!\!\!\!\!\!\!\!\!\!\!\!\!\!\!\!\!\!\!\!\!\!\!\!\!\!
\equiv 
\frac{\tilde{Z}^{\partial}_{1}\,\tilde{Z}^{\partial}_{4}}
     {\tilde{Z}^{\partial}_{2}\,\tilde{Z}^{\partial}_{3}}. 
\nonumber
\eea
\end{widetext}

For $\beta > \beta^{\ }_{c}$, the dual Ising models are in the disordered 
phase and one can perform a high-temperature expansion to calculate the 
ratio 
$
(\tilde{Z}^{\partial}_{1}\,\tilde{Z}^{\partial}_{4})
/
(\tilde{Z}^{\partial}_{2}\,\tilde{Z}^{\partial}_{3})
$. 
Using the same loop description as for the original system, with 
$t$ replaced by $\tilde{t} = \tanh(\tilde{\beta})$, one can show that 
$
(\tilde{Z}^{\partial}_{1}\,\tilde{Z}^{\partial}_{4}) 
/ 
(\tilde{Z}^{\partial}_{2}\,\tilde{Z}^{\partial}_{3}) 
= 
1
$ 
in the thermodynamic limit. 

The remaining terms in Eq.~(\ref{eq: Stopo above beta_c}) can be 
dealt with more promptly by reverting back to the original $\sigma$ spin 
degrees of freedom, 
\bea
\sum^{s,s^{\prime}_{\ } \in \delta^{\ }_{1}+\delta^{\ }_{4}}
               _{\langle s,s^{\prime}_{\ } \rangle} 
    \theta^{\ }_{s}(g) \theta^{\ }_{s^{\prime}_{\ }}(g)
\; - 
\sum^{s,s^{\prime}_{\ } \in \delta^{\ }_{2}+\delta^{\ }_{3}}
               _{\langle s,s^{\prime}_{\ } \rangle} 
    \theta^{\ }_{s}(g) \theta^{\ }_{s^{\prime}_{\ }}(g)
&=& 
\nonumber \\ 
&&
\!\!\!\!\!\!\!\!\!\!\!\!\!\!\!\!\!\!\!\!\!\!\!\!\!\!
\!\!\!\!\!\!\!\!\!\!\!\!\!\!\!\!\!\!\!\!\!\!\!\!\!\!
\!\!\!\!\!\!\!\!\!\!\!\!\!\!\!\!
= 
\sum^{\ }_{i \in \delta^{\ }_{1}+\delta^{\ }_{4}} 
    \sigma^{\ }_{i}(g) 
- 
\sum^{\ }_{i  \in \delta^{\ }_{2}+\delta^{\ }_{3}} 
    \sigma^{\ }_{i}(g) 
, 
\nonumber 
\eea
where $i$ labels the bonds of the square lattice, and $i \in \delta$ 
means that the bond $i$ connects two sites $s$ and $s^{\prime}_{\ }$ 
belonging to $\delta$. 
This contribution can be shown to vanish identically since the set of 
boundary $\sigma$ spins in bipartitions 1 and 4 is identical to the set 
of boundary spins in bipartitions 2 and 3 
(see Fig.~(\ref{fig: topological partitions})). 

In the end we find that 
\bea
S^{\ }_{\textrm{topo}} 
&=&
\lim^{\ }_{r, R \to \infty} 
\left\{\vphantom{\sum^{\ }_{g\in G}} 
\frac{1}{Z} 
  \sum^{\ }_{g \in G} e^{\beta \sum^{\ }_{i} \sigma^{\textrm{z}}_{i}(g)} 
\log^{\ }_{2} 
\frac{\tilde{Z}^{\partial}_{1} \, \tilde{Z}^{\partial}_{4}}
     {\tilde{Z}^{\partial}_{2} \, \tilde{Z}^{\partial}_{3}} 
\vphantom{\sum^{\ }_{g\in G}}\right\} 
= 
0 
\nonumber
\eea
identically in the ordered phase $\beta > \beta^{\ }_{c}$. 
%
%

\subsection{\label{sec: beyond 1-body potentials}
Beyond $1$-body potentials
           } 
As we already  mentioned, the calculations carried out in 
Sec.~\ref{sec: 1-body potentials} for the specific model presented in 
this paper can be straightforwardly extended to the case of any 
factorizable wavefunction. 
All one needs to do is identify a proper set of \emph{local} generators 
(i.e., acting on the $\sigma$ spins contained within a disc of finite radius) 
for the group $G$, and the equivalent of the 
collective boundary flip operators. 
The rest of the derivation follows essentially unchanged, in the limit 
$r,R \to \infty$. 

What happens if we attempt to generalize our approach further and we 
consider non-factorizable wavefunctions? 
For simplicity, take once again the Kitaev-like GS wavefunction in 
Eq.~(\ref{eq: Kitaev field GS}), but replace the argument of the 
exponential $\beta \sum^{\ }_{i} \sigma^{\textrm{z}}_{i}(g,\alpha) / 2$ 
with some generic function $-\beta E^{\ }_{g} / 2$. 
As we can see immediately from Eq.~(\ref{eq: rho_A}), our approach to 
compute the topological entropy can no longer be used from the very first 
stage. 
On the other hand, it is tempting to conjecture that, so long as 
$E^{\ }_{g}$ is \emph{short ranged} (i.e., it can be written as the 
sum of terms involving $\sigma$ spins within a disc of finite radius on the 
lattice), the error that one makes by neglecting the terms involving spins 
across the boundary of a bipartition ($E^{\partial}_{g}$) 
does not give topological 
contributions to $S^{\ }_{\textrm{topo}}$. 
Under this assumption, one can then set $E^{\partial}_{g} = 0$ and use 
the approximate equality 
$E^{\ }_{g} \simeq E^{A}_{g^{\ }_{A}} + E^{B}_{g^{\ }_{B}}$ 
to re-establish the factorability needed to carry on with the 
calculations. 
The result obtained for $S^{\ }_{\textrm{topo}}$ in 
Eq.~(\ref{eq: formula for the topo entropy}), 
which employed this approximation, nonetheless
shows no explicit dependence on it 
in the final expression, and one could then 
reinstate the full $E^{\ }_{g}$ at that stage.
If the conjecture above is correct, the formula in 
Eq.~(\ref{eq: formula for the topo entropy}) gives the \emph{exact} 
topological entropy for a generic GS wavefunction that satisfies 
(i) the positive amplitude condition, 
(ii) the group condition for $G$,~\cite{footnote: group condition} 
and 
(iii) the \emph{locality} (i.e., short ranged) condition on $E^{\ }_{g}$. 

A simple example where this conjecture can be applied rather 
straightforwardly is the case where Eqs.~(\ref{eq: Kitaev field Ham1}) 
and~(\ref{eq: Kitaev field GS}) are replaced by 
\bea
H 
&=& 
H^{\ }_{\textrm{Kitaev}} 
+ 
\lambda^{\ }_{1}
\sum^{\ }_{s} 
  e^{-\beta \sum^{\ }_{i \in s} \sum^{\ }_{\langle i j \rangle, j \notin s} 
     \hat{\sigma}^{\textrm{z}}_{i} \hat{\sigma}^{\textrm{z}}_{j} }_{\ } 
\label{eq: Kitaev field Ham nn}
\\ 
\vert GS \rangle 
&=& 
\frac{1}{\sqrt{Z}}
\sum^{\ }_{g \in G} 
  e^{\beta \sum^{\ }_{\langle i j \rangle} 
           \sigma^{\textrm{z}}_{i}(g) \sigma^{\textrm{z}}_{j}(g) / 2}
    g \, \vert 0 \rangle. 
\label{eq: Kitaev field GS nn}
\eea
(For a discussion of the general construction scheme of such type of 
Hamiltonians, see Ref.~\onlinecite{Castelnovo2005}.) 
Here $\vert GS \rangle$ is the GS wavefunction of $H$ in the topological 
sector where $\prod^{\ }_{i \in p} \hat{\sigma}^{\textrm{z}}_{i}=+1$. 
The notation $\langle i j \rangle, j \notin s$ stands for $j$ 
nearest-neighbor of $i$ but \emph{not} adjacent to the same vertex $s$. 
Without loss of generality, we consider the range $\beta \in (0,\infty)$, 
where the new term in the Hamiltonian favors ferromagnetic order in the 
$\sigma$ spins. 

Let us then introduce the same description in terms of the $\theta$ 
spins, as in the previous section. Given that the product of two 
nearest-neighboring $\sigma$ spins 
translates into the product of two \emph{next}-nearest-neighboring 
$\theta$ spins, we obtain 
\beq
\sum^{\ }_{g \in G} 
  e^{\beta \sum^{\ }_{\langle i j \rangle} 
     \sigma^{\textrm{z}}_{i}(g) \sigma^{\textrm{z}}_{j}(g)} 
\equiv 
\frac{1}{2} \sum^{\ }_{\bftheta \in \Theta} 
  e^{2 \beta \sum^{\ }_{\langle\langle s,s^{\prime}_{\ } \rangle\rangle} 
     \theta^{\ }_{s} \theta^{\ }_{s^{\prime}_{\ }}} 
\nonumber 
\eeq
(notice the additional factor of $2$ in the exponent due to the fact that 
the same product $\theta^{\ }_{s} \theta^{\ }_{s^{\prime}_{\ }}$ 
corresponds to two distinct products 
$\sigma^{\textrm{z}}_{i} \sigma^{\textrm{z}}_{j}$), and 
\beq
\vert GS \rangle 
= 
  \sum^{\ }_{\bftheta \in \Theta} 
    \frac{e^{\beta \sum^{\ }_{\langle\langle s,s^{\prime}_{\ } \rangle\rangle} 
          \theta^{\ }_{s} \theta^{\ }_{s^{\prime}_{\ }}}}
         {\sqrt{Z}} 
      g(\bftheta) 
\, \vert 0\rangle, 
\eeq
where 
$
Z 
= 
\sum^{\ }_{\bftheta \in \Theta} 
  e^{2 \beta \sum^{\ }_{\langle\langle s,s^{\prime}_{\ } \rangle\rangle} 
     \theta^{\ }_{s} \theta^{\ }_{s^{\prime}_{\ }}}
$. 
The latter is the partition function of a square-lattice Ising model with sole 
next-nearest-neighbor interactions, which factorizes into 
the product of two \emph{decoupled} Ising models with nearest-neighbor 
interactions (namely corresponding to the $\theta$ spins on each of the 
two sublattices). 
In this case, all equal-time correlators in the GS of the quantum system can 
be written in terms of classical correlators of two decoupled Ising models. 
As before, we expect the system to undergo a phase transition when the two 
Ising models become critical at 
$\beta^{\ }_{c} = (1/4) \ln (\sqrt{2} + 1) \simeq 0.2203434$. 
However, contrarily to the previous case, the new model undergoes a 
spontaneous symmetry-breaking (Landau-Ginzburg) phase transition! 
This is best seen by mapping the system onto a quantum eight-vertex model, 
as discussed below. 
The local order parameter that captures the transition is the 
magnetization of the $\sigma$ spins, whose expectation value can be written 
as an ensemble average of the product of two neighboring $\theta$ spins, 
i.e., belonging to two \emph{decoupled} Ising models. 
Clearly such average vanishes identically in the high-temperature phase 
$\beta < \beta^{\ }_{c}$, while it becomes finite in the ordered phase 
$\beta > \beta^{\ }_{c}$. Notice that this {\it local} order parameter that 
acquires an expectation value does so in the {\it non-topologically ordered} 
phase, as expected from the fact that no local order parameter exists that 
resolves the topological phase.

What is the fate of the topological entropy across this Landau-Ginzburg 
phase transition? 
According to the conjecture above, we can directly substitute 
$
\sum^{\ }_{i} \sigma^{\textrm{z}}_{i} 
\to 
\sum^{\ }_{\langle i j \rangle} 
           \sigma^{\textrm{z}}_{i} \sigma^{\textrm{z}}_{j}
$ 
into Eq.~(\ref{eq: formula for the topo entropy}) and compute 
$S^{\ }_{\textrm{topo}}$. 
This amounts to replacing the boundary Ising partition functions in the 
argument of the logarithm with the partition functions of two decoupled 
boundary Ising models. 
As a result, all calculations carried out in the previous section remain 
essentially unchanged and one arrives to the identical result that 
$S^{\ }_{\textrm{topo}} = 2$ throughout the high-temperature phase, and 
vanishes otherwise. 

\begin{figure*}[!ht]
\vspace{0.2 cm}
\includegraphics[width=1.98\columnwidth]{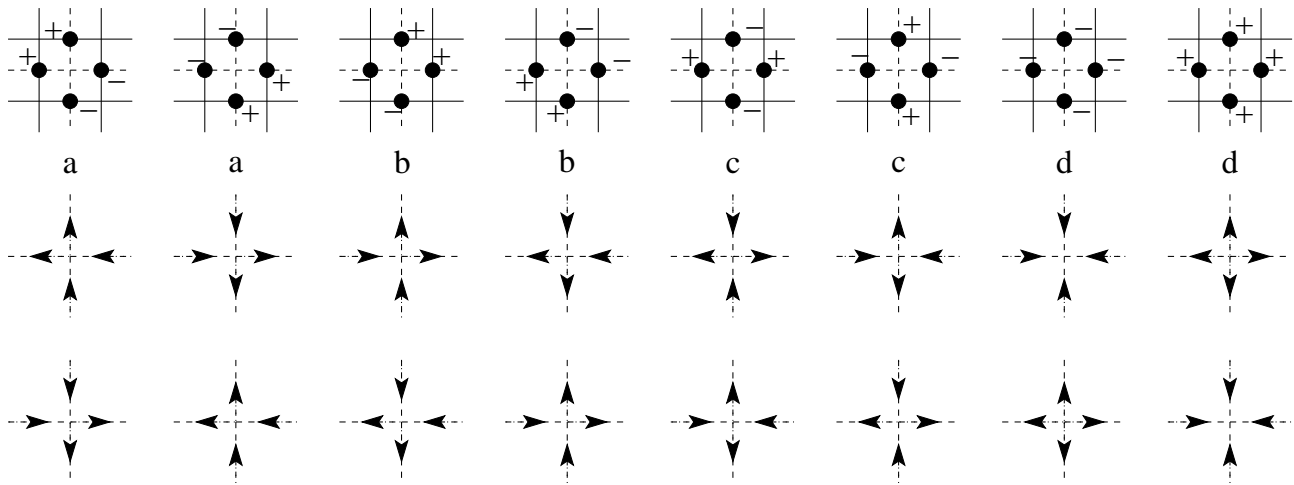}
\caption{
\label{fig: 8v mapping}
Illustration of the eight allowed spin configurations around a plaquette of 
the square lattice, in any basis state $g \vert 0 \rangle$, $g \in G$. 
These can in turn be mapped onto the configurations of an eight vertex model 
by replacing each positive spin with an arrow along the corresponding bond 
of the dual lattice (dashed lines), pointing, say,  from sublattice $A$ to 
sublattice $B$ (vice versa for the negative spins). 
Clearly, the corresponding eight vertex configurations differ depending on 
the location of the plaquette in the dual lattice: sublattice $A$ (middle), 
or sublattice $B$ (bottom). 
The letters $a,b,c,d$ correspond to the usual labeling of the vertex 
fugacities in the eight vertex model. 
}
\end{figure*}

This scenario is in agreement with previous results on a quantum version of 
the eight-vertex model by Ardonne~\textit{et al.},~\cite{Ardonne2004} whose 
GS is a generalization of the one in our model. 
Consider indeed the wavefunction in Eq.~(\ref{eq: Kitaev field GS nn}). 
Given the nature of the group $G$, the four spins belonging to any plaquette 
of the square lattice can assume only eight distinct configurations 
($\prod^{\ }_{i \in p} \hat{\sigma}^{\textrm{z}}_{i}=+1$), illustrated in 
Fig.~\ref{fig: 8v mapping}. 
Such configurations map naturally onto the vertices of an eight-vertex model 
upon replacing each positive spin with an arrow along the corresponding bond 
of the dual lattice, pointing, say,  from sublattice $A$ to 
sublattice $B$, and vice versa for the negative spins (as shown 
in Fig.~\ref{fig: 8v mapping}). 
Given that 
$
\sum^{\ }_{\langle i j \rangle} \cdots 
\equiv 
\sum^{\ }_{s} \sum^{\ }_{\langle i j \rangle \in s} \cdots
$, 
the amplitudes in the GS wavefunction~(\ref{eq: Kitaev field GS nn}) factorize 
into products of vertex fugacities 
$
\exp(\beta \sum^{\ }_{\langle i j \rangle \in s} 
     \sigma^{\textrm{z}}_{i} \sigma^{\textrm{z}}_{j} / 2)
$. 
In the notation of Fig.~\ref{fig: 8v mapping}, the vertex fugacities assume 
the values $a = b = 1$, $c = e^{-2 \beta}_{\ }$ and $d = e^{2 \beta}_{\ }$. 
The GS spatial properties of our model are therefore captured by a classical 
eight-vertex model with the appropriate 
fugacities~\cite{Ardonne2004} (but see Ref.~\onlinecite{Castelnovo2005} 
for a general discussion of such quantum-to-classical correspondence), and 
one can then use Baxter's exact solution~\cite{Baxter_book} to obtain the 
phase diagram as well as the scaling exponents at the critical point. 
All this is discussed in detail in Ref.~\onlinecite{Ardonne2004}: 
the model undergoes a second-order, $\mathbb{Z}^{\ }_{2}$-symmetry-breaking 
phase transition when $d^{2}_{\ } = c^{2}_{\ } + 2$ 
(i.e., $\beta^{\ }_{c} = (1/4) \ln (\sqrt{2} + 1)$), separating a 
topologically ordered liquid phase from a Landau-Ginzburg ordered phase. 
The local order parameter across the transition is indeed the magnetization 
in the original $\sigma$ spins. 
%
%
%

\section{\label{sec: concl}
Conclusions
        } 
In this paper we studied a topological quantum phase transition in a
microscopic model that can be examined analytically. For this system,
an extension of the toric code, the ground state wavefunction can be
written exactly as a function of the parameter $\beta$ that drives the
system across the quantum phase transition. We computed the
topological entropy for this system as a function of $\beta$, and
showed that it remains at a constant non-zero value throughout the
topologically ordered phase ($\beta < \beta^{\ }_{c} \simeq
0.4406868$). Immediately after the quantum phase transition at
$\beta_c$, the topological entropy drops to zero and remains so in the
non-topologically ordered phase ($\beta > \beta^{\ }_{c}$).

The GS wavefunction of our quantum system has positive amplitudes in
the basis of choice. This property allows us to relate many
quantities that are relevant in characterizing the (2+1 D) quantum
system to those of a simple (2D, not 3D) classical Ising model at an
inverse (classical) temperature equal to the value of the coupling
constant $\beta$ that drives the quantum system through the $T=0$ phase
transition. For example, the magnetization of the quantum system
equals the energy $E_{\rm Ising}(\beta)$ of the classical Ising
model. While the magnetization is continuous and non-vanishing across the 
quantum phase transition (much as the energy $E_{\rm Ising}(\beta)$ is 
across the classical Ising transition), its derivative with respect to 
$\beta$ diverges logarithmically at $\beta_c$ (much as the Ising model 
heat capacity $C_{\rm Ising}$ diverges logarithmically at $\beta_c$). 

Despite the relation to the 2D classical Ising model, the quantum
phase transition {\it does not} have a local order parameter that
vanishes on one side and not on the other. Of course one expects that
no local order parameter can characterize the topological phase, but
in this particular example, there is no order parameter that characterizes 
the non-topologically ordered phase either. One can indeed identify from
the mapping to the Ising model a parameter that orders in the
non-topological phase; however, this variable is {\it non-local} in
the physical spin variables used to define the local
Hamiltonian. Specifically, the order parameter is, in the language used 
in this paper, the expectation value of the $\theta_s$ variables defined on 
the sites of the square lattice, such that $\sigma_{\langle
ss'\rangle}=\theta_s\theta_s'$, for nearest neighboring sites
$s,s'$. While $\sigma_{\langle ss'\rangle}$ is obviously local in
terms of the $\theta_s,\theta_s'$, the inversion needed to write the
$\theta$'s in terms of the $\sigma$'s is non-local. Hence,
$\langle\theta_s\rangle$ may detect the transition into the
non-topological phase, but as it is non-local it is not an order
parameter in the usual sense. That there is no order parameter for the
non-topologically ordered phase is not generic (see the example in
Sec.~\ref{sec: beyond 1-body potentials}), as perhaps the most obvious
exit from a topological phase is by escaping into a locally ordered phase 
due to spontaneous symmetry breaking. 
Hence, the main example studied in this paper is
particularly interesting in that one has {\it no local order parameter
in either phases}.

Recently, P. Zanardi {\it et al.} proposed a new approach to study 
quantum phase transitions through the behavior of a metric 
tensor in parameter space, derived from a fidelity-based notion of distance 
between states.~\cite{Zanardi2007} 
In this approach, one  does not require any {\it a priori} 
knowledge of an order parameter to detect a phase transition 
(one may argue that the fidelity encompasses correlations of both local and 
non-local operators). Thus, these ideas may be particularly useful to 
detect topological quantum phase transitions (see 
Ref.~\onlinecite{Hamma2007}). 
 
We end with a speculative note: topological quantum phase transitions 
should share the feature that, even in the case when there is no 
local order parameter in either the topological or the 
non-topological phase, there should be, generically, detectable 
singularities in high enough derivatives of local observables with respect 
to the coupling that takes the system across the transition 
(as in the case study presented here).~\cite{footnote: Zohar} After all, 
what would be a phase transition without singularities in any 
physical observable? 
 
Upon completion of this work, we became aware of similar work being pursued 
from a gauge theoretical perspective by S.~Papanikolaou, 
K.~Raman and E.~Fradkin on the quantum eight-vertex 
model,~\cite{Papanikolaou2007} to which we refer the reader for a 
complementary approach. 

\section*{
Acknowledgments
         }
We would like to thank Xiao-Gang Wen, Eduardo Fradkin and Stefanos 
Papanikolaou for enlightening discussions. We are particularly indebted to 
Paul Fendley, whose comments brought us to consider the model discussed in 
Sec.~\ref{sec: beyond 1-body potentials}. 
This work is supported in part by the NSF Grants DMR-0305482 and
DMR-0403997 (CC and CC), 
and by EPSRC Grant No. GR/R83712/01 (C.~Castelnovo). 
C.~Castelnovo would like to acknowledge the I2CAM NSF Grant DMR No. 0645461 
for travel support, during which part of this work was carried out. 
%
%

\end{document}